\documentclass[sigconf,natbib=true,review=False,anonymous=False]{acmart}
\usepackage{multirow}
\usepackage{booktabs}
\usepackage{adjustbox}
\usepackage{amsmath}
\usepackage{pifont}
\usepackage{enumitem}
\usepackage{lipsum}
\setlist[itemize]{leftmargin=*}
\def\BibTeX{{\rm B\kern-.05em{\sc i\kern-.025em b}\kern-.08em
    T\kern-.1667em\lower.7ex\hbox{E}\kern-.125emX}}
\AtBeginDocument{%
  \providecommand\BibTeX{{%
    \normalfont B\kern-0.5em{\scshape i\kern-0.25em b}\kern-0.8em\TeX}}}

\setcopyright{acmcopyright}
\copyrightyear{2018}
\acmYear{2018}
\acmDOI{XXXXXXX.XXXXXXX}

\acmConference[SIGIR'23]{Make sure to enter the correct
  conference title from your rights confirmation email}{July 23-27,
  2023}{Taipei, Taiwan}
\acmBooktitle{SIGIR '23: ACM Symposium on Neural Gaze Detection, July 23-27, 2023, Taipei, Taiwan} 
\acmPrice{15.00}
\acmISBN{978-1-4503-XXXX-X/18/06}

\begin{document}
\copyrightyear{2023}
\acmYear{2023}
\setcopyright{acmlicensed}\acmConference[SIGIR '23]{Proceedings of the 46th International ACM SIGIR Conference on Research and Development in Information Retrieval}{July 23--27, 2023}{Taipei, Taiwan}
\acmBooktitle{Proceedings of the 46th International ACM SIGIR Conference on Research and Development in Information Retrieval (SIGIR '23), July 23--27, 2023, Taipei, Taiwan}
\acmPrice{15.00}
\acmDOI{10.1145/3539618.3591689}
\acmISBN{978-1-4503-9408-6/23/07}
%%
%% The "title" command has an optional parameter,
%% allowing the author to define a "short title" to be used in page headers.
% \title{MFIRec: Multi-Frequency Interest circular convolution Network for sequential recommendation}
\title{Frequency Enhanced Hybrid Attention Network for Sequential Recommendation}
%%
%% The "author" command and its associated commands are used to define
%% the authors and their affiliations.
%% Of note is the shared affiliation of the first two authors, and the
%% "authornote" and "authornotemark" commands
%% used to denote shared contribution to the research.
\author{Xinyu Du}
\authornote{They are co-first authors with equal contribution.}
\email{xydu@stu.suda.edu.cn}
\affiliation{%
  \institution{Soochow University}
  \streetaddress{P.O. Box 1212}
  \city{Suzhou}
  \state{Jiangsu}
  \country{China}
  \postcode{43017-6221}
}

\author{Huanhuan Yuan}
\authornotemark[1]
% \authornote{They are co-first authors with equal contributions.}
\email{hhyuan@stu.suda.edu.cn}
\affiliation{%
  \institution{Soochow University}
  \streetaddress{P.O. Box 1212}
  \city{Suzhou}
  \state{Jiangsu}
  \country{China}
  \postcode{43017-6221}
}

\author{Pengpeng Zhao}
\authornote{Corresponding author.}
\email{ppzhao@suda.edu.cn}
\affiliation{%
  \institution{Soochow University}
  \streetaddress{P.O. Box 1212}
  \city{Suzhou}
  \state{Jiangsu}
  \country{China}
  \postcode{43017-6221}
}

\author{Jianfeng Qu}
\email{jfq@suda.edu.cn}
\affiliation{%
  \institution{Soochow University}
  \streetaddress{P.O. Box 1212}
  \city{Suzhou}
  \state{Jiangsu}
  \country{China}
  \postcode{43017-6221}
}
% Artificial Intelligence
\author{Fuzhen Zhuang}
\email{zhuangfuzhen@buaa.edu.cn}
\affiliation{%
  \institution{Institute of Artificial Intelligence \& SKLSDE, School of Computer Science, Beihang University}
  \streetaddress{P.O. Box 1212}
  \city{Beijing}
  % \state{Jiangsu}
  \country{China}
  \postcode{43017-6221}
}

\author{Guanfeng Liu}
\email{guanfeng.liu@mq.edu.au}
\affiliation{%
  \institution{Department of Computing, Macquarie University}
  \streetaddress{P.O. Box 1212}
  \city{Sydney}
  \country{Australia}
  \postcode{43017-6221}
}

% \author{Yanchi Liu}
% \email{yanchi.liu@rutgers.edu}
% \affiliation{%
%   \institution{Rutgers University}
%   \streetaddress{P.O. Box 1212}
%   \city{New Brunswick}
%   \state{New Jersey}
%   \country{USA}
%   \postcode{43017-6221}
% }

\author{Victor S. Sheng}
\email{victor.sheng@ttu.edu}
\affiliation{%
  \institution{Texas Tech University}
  \streetaddress{P.O. Box 1212}
  \city{Lubbock}
  \state{Texas}
  \country{USA}
  \postcode{43017-6221}
}

\renewcommand{\shortauthors}{Xinyu Du et al.}

%%
%% The abstract is a short summary of the work to be presented in the
%% article.
\begin{abstract}
% 在建模用户兴趣和偏好的同时，我们认为序列推荐是一个序列预测问题。
% 从item-embedding压缩后的波动图可以看出，用户的行为具有一定的周期性，用户会点击特征相同或者相关的物品，用户的兴趣也在随之发生变化。
% 序列推荐时做什么的
% 进入话题self-attention
% 最近工作，现有工作主要的关注点是什么。
% (1) 第一点【Hybrid self-attention】
% A：现有工作都是在时域，计算复杂度高，只能捕获item-level的。
% B: 我们提出频域self-attention。

% (2) 第二点【FMP-Rec使用全部频率分量，FEDformer均匀分布采样】
% A：现有的再频域中的工作往往使用【全部的频率components】，全频段，或者使用均匀分布采样。
% B：我们提出两种频率采样策略，并且使用频率渐变结构，可以有效减少每层的计算量。

% (3) 第三点【频率loss，借鉴CV】
% A：现有的正则和损失函数都是【只在时域】中约束模型的训练。
% B：我们提出频域loss
% 综上
% 我们时第一个再序列推荐中使用频域和时域双attention机制，并且提出的frequency component采样策略能够再保证有效提取特征的同时，大幅减小参数量。通过使用频域和时域dual domain regularization可以很好约束模型的训练过程并得到最终的表示。
The self-attention mechanism, which equips with a strong capability of modeling long-range dependencies, is one of the extensively used techniques in the sequential recommendation field. However, many recent studies represent that current self-attention based models are low-pass filters and are inadequate to capture high-frequency information. 
Furthermore, since the items in the user behaviors are intertwined with each other, these models are incomplete to distinguish the inherent periodicity obscured in the time domain. 
% That makes these models more efficient to learn user global-term preferences but overlooks the local-term dynamics.
In this work, we shift the perspective to the frequency domain, and propose a novel \textbf{F}requency \textbf{E}nhanced Hybrid \textbf{A}ttention Network for Sequential \textbf{Rec}ommendation, namely \textbf{FEARec}. 
In this model, we firstly improve the original time domain self-attention in the frequency domain with a ramp structure to make both low-frequency and high-frequency information could be explicitly learned in our approach.
Moreover, we additionally design a similar attention mechanism via auto-correlation in the frequency domain to capture the periodic characteristics and fuse the time and frequency level attention in a union model.
Finally, both contrastive learning and frequency regularization are utilized to ensure that multiple views are aligned in both the time domain and frequency domain.
Extensive experiments conducted on four widely used benchmark datasets demonstrate that the proposed model performs significantly better than the state-of-the-art approaches\footnote{Our code is available at: https://github.com/sudaada/FEARec.}. 
% Our code is available at: https://anonymous.4open.science/r/xxx.

\end{abstract}

%%
%% The code below is generated by the tool at http://dl.acm.org/ccs.cfm.
%% Please copy and paste the code instead of the example below.
%%
%\ccsdesc[300]{Information systems~Collaborative filtering}

% \ccsdesc[500]{Computer systems organization~Embedded systems}
% \ccsdesc[300]{Computer systems organization~Redundancy}
% \ccsdesc{Computer systems organization~Robotics}
% \ccsdesc[100]{Networks~Network reliability}

\maketitle
\section{Introduction}
\label{introduction}
% 第一段
% Sequential recommendation system aims to model the dynamic preferences in users’ historical interactions and predicts the subsequent items that users will probably interact with in the future
In recent years, sequential recommendation \cite{BERT4Rec, GRU4Rec} has attracted increasing attention from both industry and academic communities.
Essentially, the key advantage of sequential recommender models lies in the explicit modeling of item chronological correlations.
To capture users' dynamic sequential information more accurately, recent years have witnessed lots of efforts based on either Markov chains~\cite{Markov} or Recurrent Neural Networks (RNNs)~\cite{GRU4Rec}.
% 一些模型的历史
%Traditional methods [18, 36, 37] are based on the Markov Chain (MC) assumption that the next item only depends on previous items.
% With the advancements in deep learning in the recent past, various models employ deep neural networks, such as Convolutional Neural Networks (CNNs) [40] and Recurrent Neural Networks (RNNs) [19, 20], as base sequence encoders to generate hidden representations of sequences. 
% The limitations are that CNNs are only effective in capturing local features [40] while RNNs display poor parallelism capacity [2].
Meanwhile, Transformer~\cite{Attention} has emerged as a powerful architecture and a dominant force in various research fields and outperformed RNN-based models in the recommendation tasks.
Attributed to its strong capability of modeling long-range dependencies in data, various sequential recommender models~\cite{SASRec, CoSeRec, DuoRec, BERT4Rec, CL4SRec, lspp, FDSA, MMInfoRec, ICLRec} adopt Transformers as the sequence encoder to capture item correlations via assigning attention weights to items at different positions and obtain high-quality sequence representations.

% This is largely attributed to its strong capability of modeling long-range dependencies in data with the multi-head self-attention mechanism, which could automatically assign attention weights to items at different positions.
% \begin{figure}[!t]
% 	\centering
% 	{
% 	\includegraphics[width=1\linewidth]{user_item10102.pdf}}
% 	\caption{Illustration of two different attention.
% 	Vanilla Time Domain Attention (a) adapts the fully connection among all items. Periodic Frequency Domain Attention (b) focuses on the connections of sub-series among potential periods.
% 	}
% 	\label{fig_user_item}
% 	% \vspace{-1.1em}
% \end{figure}

\begin{figure}[!t]
	\centering
	{
	\includegraphics[width=1\linewidth]{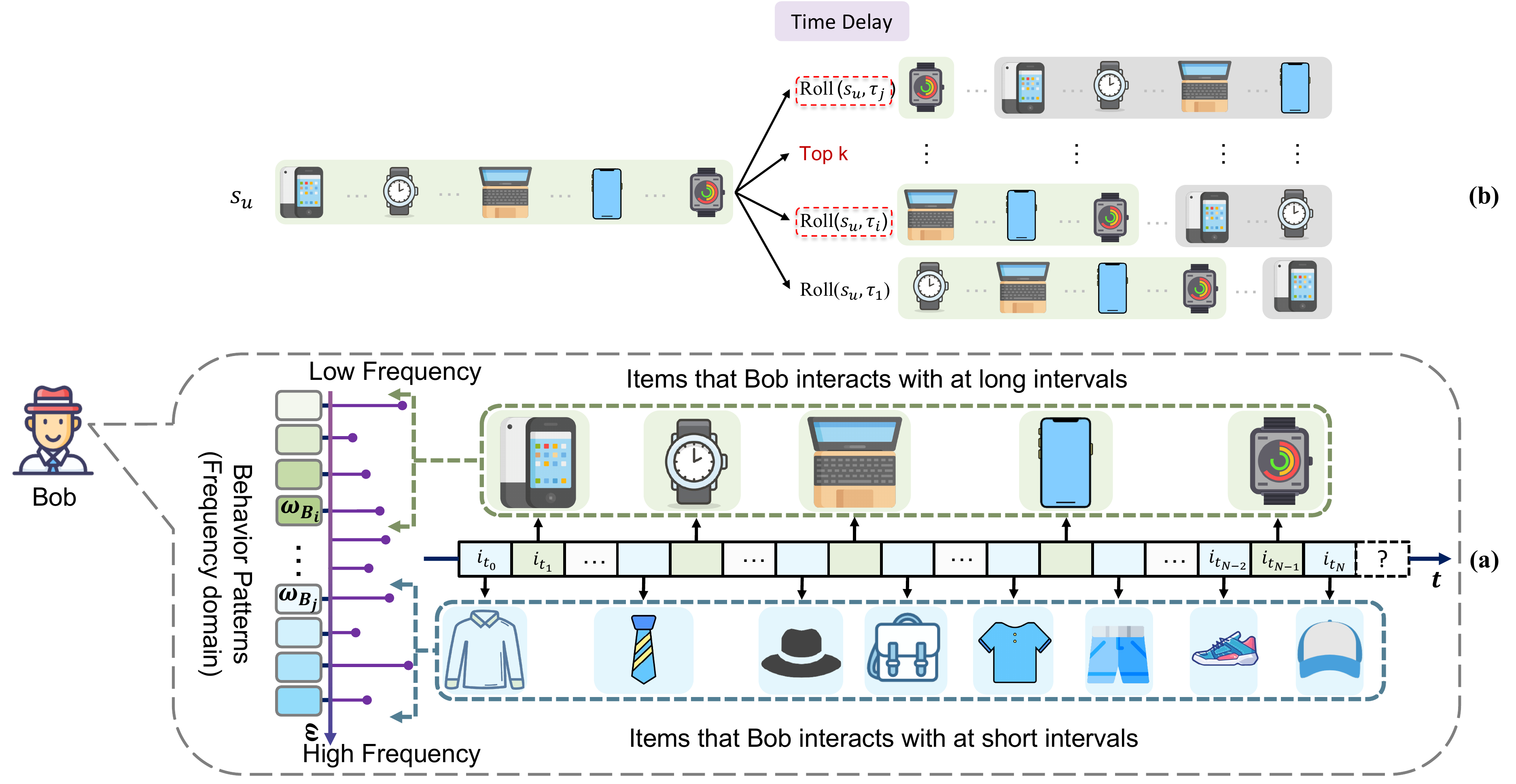}}
	\caption{Illustration of our motivations.
	(a) By shifting the sequential behaviors from the time domain to the frequency domain, the historical item sequence of each user decomposed into multiple behavioral patterns with different frequencies and periods along the $\omega$-axis.
	(b) By constructing $s_u$'s $\tau$ lag sequences $\operatorname{Roll}(s_u,\tau)$, the periodic pattern can be distinguished from the most similar sequences with the help of discrete Fourier transform.}
	\label{fig_user_item}
	% \vspace{-1.1em}
\end{figure}

% 第二段
Despite their effectiveness, self-attention used in current Transformer based models is constantly a low-pass filter, which continuously erases high-frequency information according to the theoretical justification in~\cite{howVITwork, VIT_seelike_CNN}.
% Current Transformer-based models suffer
% from sufficiently capturing the importance of short-term user dynamics, owing to the fact that it is a natural non-local operation~\cite{LOCKER}.
That means, SASRec \cite{SASRec} and its variants are highly capable of capturing low-frequencies in the sequential data~\cite{iFormer}, which helps capture a global view of user’s interaction data and the overall evolution of user preference. 
But owing to its non-local operation~\cite{LOCKER}, they are incompetent in learning high-frequencies information, including those items that interact frequently within a short period.

% This can be intuitively explained: self-attention is a global operation and much more capable of capturing global information (low frequencies) in the data than local information (high frequencies) \cite{VIT_CNN, iFormer}.
% In other words, self-attention in terms of its spectral-domain effect is constantly a low-pass filter, which continuously erases high-frequency information according to the theoretical justification in \cite{lowpassfilter}.
% 为了改变self-attention缺乏的local的特点。
To alleviate these issues, existing methods import local constraints in different ways to complement Transformer-based models. 
Such as LSAN \cite{LSAN} adopts a novel twin-attention paradigm to capture the global and local preference signals via a self-attention branch and a convolution branch module, respectively.
L{\scriptsize{OCKER}} \cite{LOCKER} combines five local encoders with existing global attention heads to enhance short-term user dynamics modeling.
% to harmonize different frequency component 
% 文章的核心部分！！！
% 然而这些模型都是从时域选择使用不同的模块来提取global和local特征。
However, the above-mentioned models almost process the historical interactions from the perspective of the time domain, but seldom consider tackling this challenge in the frequency domain, 
% \textcolor{blue}{
where the reduced high-frequency information can be easily obtained.
% }
%  which is commonly used in the digital signal processing area to filter noise.
% (过渡到频域的概念)提到频域了，必须得介绍频域，把频域的概念弄清楚。
Take Figure~\ref{fig_user_item}(a) for an example, from the viewpoint of the time domain, all the items are chronologically ordered and intertwined along the $t$-axis.
With a Discrete Fourier Transform (DFT)~\cite{FMLP-Rec, GFNet}, the historical item sequence features of each user can be decomposed into multiple behavioral patterns with different frequencies along the $\omega$-axis, and the input time features are then converted to the frequency domain. 
Eventually, improving the self-attention operation's capability for capturing high-frequency information in the frequency domain could become a more direct and effective way. 
% 第二点：为什么要使用频域attention以及结合的好处。
% 举例子
% 存在用户存在一种潜在intent dependicy， hide 在复杂的时间序列里，直接在时域里做，不好直接发现其周期性（原因，频域上做更好），举例子，123插一个4，（这就是为什么不好）（频域里为什么好，频域做周期和attention好），说是怎么解决的（），在频域里有先天性的优势天然存在的特征（引一篇文献），维纳-辛磬定律，子序列频率周期性是否相同。
% 识别这个东西，可以通过这种方法，但是不好做，time series中已经做了。做完了有什么好处，为了在推荐系统中捕获也不活。
% 我们的方法是
% (怎么引入频域attention，频域是频域的好处)
% Moreover, intricate temporal patterns of the long-term item interaction prohibit the model from finding reliable dependencies when adopting the time domain self-attention mechanism.
% 前面说的是改进频域attention。
% 存在用户存在一种潜在intent dependicy，hidden在复杂的时间序列里.

% transformer: strong capability of modeling long-range dependencies.(但是是低频，所以有改进，但没有考虑频域)
% \textcolor{blue}{
Moreover, users’ behaviors on the Internet tend to show certain periodic trends \cite{merit_11, merit_19, merit_21}.
% }
Take Figure \ref{fig_user_item}(b) for an example, 
% As shown in Figure \ref{fig_user_item}(b), 
there is a periodic behavior pattern hidden in Bob's low-frequency interactive item, $i.e.$, a watch and laptop are brought after buying a mobile phone. When he recently bought a product with similar characteristics again, a laptop may be a good recommendation to him.
% When he buys a mobile phone, he will buy a watch and a laptop. 
% 时域attention很难发现。
However, it is difficult to find the periodic behavior patterns hidden in the sequence by directly calculating the overall attention scores of items in the time domain.
% constructing models in the frequency domain help to discover inherent dependencies obscured by entangled and intricate temporal patterns.
But in the frequency domain, there emerges some methods~\cite{autoformer} constructing models to recognize the periodic characterize with the help of the Fourier transform, which inspires us to tackle this challenge from a new perspective for recommendation. 
% But with the help of the Fourier transform, the auto-correlation of different time delay sequences of the current user sequence can be efficiently obtained~\cite{autoformer}, and thus recognize the periodic characterize by aggregating the most relative delay sequence. 
% However, complex behavior patterns of users and noisy items prevent self-attention based models from directly finding these similar temporal process hidden in chronological sequences of items in time domain.
% 如图所示：
% Therefore, we try to construct a sub-sequence level connection based on the process similarity derived by user behavior periodicity.
% By calculating the auto-correlation of different delay sequences of the current user sequence, the correlation between the user's recent and past subsequences can be obtained . 
% The delay subsequence with the highest auto-correlation score implies the periodic behavior of the user.
% 直接在时域里做，不好直接发现其周期性（原因，频域上做更好），
% 举例子，123插一个4，（这就是为什么不好）（频域里为什么好，频域做周期和attention好），说是怎么解决的（）
% However, complex behavior patterns of users and noisy items prevent self-attention based models from directly finding these similar temporal process hidden in chronological sequences of items in time domain.
% 特定周期/频率的叫做行为模式
% 1-2-3的也叫做行为模式（用户的潜在的兴趣和意图，兴趣点）
% 在频域里有先天性的优势天然存在的特征（引一篇文献），维纳-辛磬定律，子序列频率周期性是否相同。

% 第三段，第三个点（对比学习，对比正则化，以及频域loss）
% 要不要单独说，还是放在最后提出解决方法。

% 解决办法：

% 第三个点：现有的对比学习和正则化都是在时域的，我们提出频域loss。

% 最后：总结提出模型。
For these reasons, we shift the perspective to the frequency domain and propose a novel \textbf{F}requency \textbf{E}nhanced Hybrid \textbf{A}ttention Network for Sequential \textbf{Rec}ommendation (\textbf{FEARec}).
Firstly, we improve the time domain self-attention with an adaptive frequency ramp structure. 
After DFT, we select a specific frequency component as the input feature for each time domain self-attention layer.
% After DFT, we 
That enables each layer to focus on different frequency ranges (including not only low-frequency but also high-frequency) to address the problem that self-attention only concentrates on low-frequency.
% 第二点：具体说怎么做的。如何体现周期性。（autoformer的原理，别人懂，回答第二个问题，好在哪里，你是怎么做的，有什么优势）
% item：达到了，做到了，总结。
% Furthermore, to disentangle the time sequence and highlight the inherent periodicity of user behaviors
Furthermore, to disentangle the temporal sequence and highlight the inherent periodicity of user behaviors, we design a novel auto-correlation based frequency attention, which is an efficient method to discover the period-based dependencies by calculating the autocorrelation in the frequency domain based on the Wiener-Khinchin theorem~\cite{wiener}. 
Autocorrelation is used to compare a sequence with a time-delayed version of itself.
Specifically, given a user's behavior sequence $s_u$ and its $\tau$ lag sequences $\operatorname{Roll}(s_u,\tau)$ in Figure~\ref{introduction}, the sequential-level connection can be constructed by aggregating the top-$k$ relative sequences ($\operatorname{Roll}(s_u, \tau_1), \cdots, \operatorname{Roll}(s_u,\tau_k)$) based on the auto-correlation.
% We computed the autocorrelation of user sequence based on the Wiener-Khinchin theorem\cite{wiener} using the sampled frequency components from the frequency domain.
% and aggregates similar sub-series by time delay aggregation
% Compared with time domain self-attention, frequency domain attention not only, but also capture user preference in sub-sequence level. 
% 两个结合：
Then, we combine the time domain self-attention with frequency domain attention in a union hybrid attention framework.
With the help of contrastive and frequency domain regularize loss, a multi-task learning method is applied to increase the supervision information and regularize the training process.
The main contributions of this paper can be summarized as follows: 
\begin{itemize}[topsep = 5 pt]
    % 原始版本
    \item We shift the perspective to the frequency domain and design a frequency ramp structure to improve existing time domain self-attention.
    % It can effectively trade-off users’ different periodic pattern across different layers. 
    \item We propose a novel frequency domain attention based on an auto-correlation mechanism, which discovers similar period-based dependencies by aggregating most relative time delay sequences.
    \item We unify the frequency ramp structure with vanilla self-attention and frequency domain attention in one framework and design a frequency domain loss to regularize the model training.
    \item We conduct extensive experiments on four public datasets, and the experimental results imply the superiority of the FEARec compared to state-of-the-art baselines.
\end{itemize}

% model图
\begin{figure*}[!t]
	\centering
	{\includegraphics[width=0.98\linewidth]{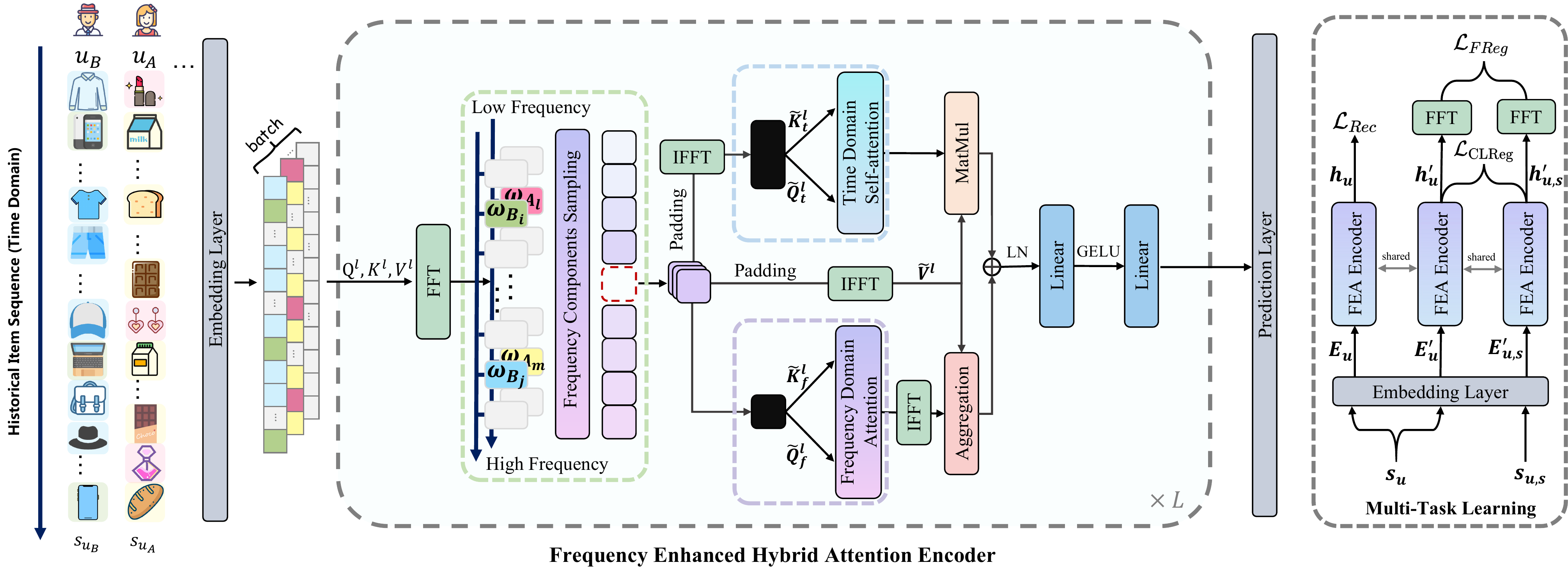}}
	\caption{An overview of FEARec. 
	FEARec generates item embedding with positional embedding through the embedding layer, then stacks $L$ hybrid attention blocks to extract user preference in both the item level and frequency level. Supervised contrastive learning and frequency domain regularization are also applied as auxiliary tasks to complement the main task.}
	% 对于multi-task的学习
    %这啥意思
    % Note that the frequency range of the learnable filter in each layer depends on the position of the layer in the neural network. 
    % For simplicity, we draw all the learnable filters of different layers in one block, which looks like the filter slides on the frequency domain feature across different layers.}
	\label{model}
\end{figure*}
\section{RELATED WORK}
\subsection{Sequential Recommendation} 
Sequential recommendation forecasts future items throughout the user sequence by modeling item transition correlations.
Early SR studies are often based on the Markov Chain assumption~\cite{Markov}.
% , RNN~\cite{GRU4Rec}.
% \textcolor{blue}{
Afterward, many deep learning-based sequential recommender models have been developed, e.g., GRU4Rec~\cite{GRU4Rec}, Caser~\cite{Caser}.
% }
% Afterward, numerous sequential recommender models powered by deep learning were created.
% GRU4Rec~\cite{GRU4Rec} is the very first attempt to utilize the GRU network in SR.
% % 下一句解开注释：
% To discover sequential patterns, Caser~\cite{Caser} employs both horizontal and vertical convolutional filters.
Later on, self-attention networks have shown great potential in modeling sequence data and a variety of related models are developed, e.g., SASRec \cite{SASRec}, BERT4Rec \cite{BERT4Rec} and S${^3}$Rec \cite{S3Rec}.
% Recently, More recently, In terms of 
Recently, many improvements on self-attention based solutions \cite{related28,related15,related7,related5,related29,related18,related3, STOSA} are proposed to learn better representations of user preference.
% More recently, FMLP-Rec \cite{FMLP-Rec} utilizes an all-MLP structure without self-attention mechanism for SR.
% In terms of the training scheme, most of these methods follow the next-item supervised training style.
% 解开注释
Most of them use the next-item supervised training style as their training scheme.
The other training scheme usually has extra auxiliary training tasks.
CL4SRec \cite{CL4SRec} applies a contrastive strategy to multiple views generated by data augmentation.
CoSeRec \cite{CoSeRec} introduces more augmentation operations to train robust sequence representations. 
DuoRec \cite{DuoRec} combines recommendation loss with unsupervised learning and supervised contrastive learning to optimize the SR models.
Despite the success of these models in SR, they ignored important information hidden in the frequency domain.
% \textcolor{blue}{
% Despite the success of these models in SR, they ignored important}

\subsection{Frequency Domain Learning} 
Fourier transform has been an important tool in digital signal processing and graph signal processing  for decades~\cite{related35, related1, SIGIR2022, graph}.
There are a variety of works that incorporate Fourier transform in computer vision \cite{learninginFD, fastfourierconvolution, resolution, FFC-SE, GFNet} and natural language processing \cite{NLP1, FNET}.
% 解开注释
% FNet \cite{FNET} resembles the MLP-mixer with token mixer simply being pre-fixed DFT.
% No filtering is done to adapt the data distribution. 
% 解开注释
% Global filter networks (GFNet) \cite{GFNet} learn Fourier filters to perform depthwise global convolution.
Very recent works try to leverage Fourier transform enhanced model for long-term series forecasting \cite{autoformer, FEDformer} and partial differential equations solving \cite{FNO, AFNO, UFNO}. 
However, there are few Fourier-related works in the sequential recommendation.
% More recently, FMLP-Rec \cite{FMLP-Rec} utilizes an all-MLP structure without self-attention mechanism for SR.
More recently, FMLP-Rec \cite{FMLP-Rec} first introduce a filter-enhanced MLP for SR, which multiplies a global filter to remove noise in the frequency domain.
However, the global filter tends to give greater weight to low frequencies and underrate relatively high frequencies.

\section{Proposed Method}
In this section, we present the details of the proposed \textbf{F}requency \textbf{E}nhanced Hybrid \textbf{A}ttention Network for Sequential \textbf{Rec}ommendat-
ion namely (\textbf{FEARec}).
As shown in Figure \ref{model}, after the embedding layer we first transform the embeddings of item sequences from the time domain to the frequency domain by using Fast Fourier Transform (FFT) algorithm mentioned in Appendix~\ref{FFT}.
% \textcolor{blue}{
Then hybrid attention is conducted on the sampled frequency components, which captures attention scores and periodic behavior patterns of different frequency bands simultaneously.
% }
% Then hybrid attention is conducted on the sampled frequency components, which captures user preferences in both time and frequency domain simultaneously.
% to capture 
% attention scores and periodic behavior patterns of different frequency bands
% , as illustrated in the left part of Figure \ref{model}.
Finally, we apply contrastive learning and frequency domain loss to improve representations in both time and frequency domains
\subsection{Embedding Layer}
Sequential Recommendation focus on modeling the user behavior sequence of implicit feedback, which is a list of item IDs in SR. 
In this paper, the item ID set is denoted by $\mathcal{I} = \{i_1, i_2, ..., i_{|\mathcal{I}|}\}$ and user ID set is represented as $\mathcal{U} = \{u_1, u_2, ..., u_{|\mathcal{U}|}\}$, where $i \in \mathcal{I}$ denotes an item and $u \in \mathcal{U}$ denotes a user. 
% The total numbers of users and items can be denoted as $|\mathcal{U}|$ and $|\mathcal{I}|$, respectively.
In this way, the set of user behavior can be represented as $S = \{s_1, s_2,...,s_{|\mathcal{U}|}\}$. In SR, the user's behavior sequence is usually chronologically ordered, $i.e.$, $s_u = [i_1^{(u)}, i_2^{(u)},..., i_t^{(u)},...,i_{N}^{(u)}]$, where $s_u \in S, u \in \mathcal{U}$, and $i_t^{(u)} \in \mathcal{I}$ is the item that user $u$ interacts at time step $t$ and $N$ is the sequence length. 
For $s_u$, it is embedded as:
\begin{equation}
    \mathbf{E}_{u}
    =[\mathbf{e}_{1}^{(u)},\mathbf{e}_{2}^{(u)} , ..., \mathbf{e}_{N}^{(u)}]
\end{equation}
where $\mathbf{e}_{t}^{(u)}$ is the embedding of item $i_{t}^{(u)}$.
To make our model sensitive to the positions of items, we adopt positional embedding to inject additional positional information while maintaining the same embedding dimensions of the item embedding. Moreover, dropout and layer normalization operations are also implemented:
\begin{equation}
    \mathbf{E}_{u}=\operatorname{Dropout}(\text {LayerNorm}(\mathbf{E}_{u}+\mathbf{P}))
\end{equation}
% \textcolor{blue}{
It is worth noting that each item has a unique ID different from each other, but after the embedding layer, similar items may have the same feature value.
% }

\subsection{Frequency Enhanced Hybrid Attention Encoder}
Based on the embedding layer, we develop the item encoder by stacking $L$ Frequency Enhanced hybrid Attention (FEA) blocks, 
which generally consists of three modules, $i.e.$, frequency ramp structure, hybrid attention layer, and the point-wise Feed Forward Network (FFN).
Since FEARec focuses on capturing specific frequency spectrum at different layers, we first introduce the frequency ramp structure for each layer, and then present other modules for the sampled frequencies at a certain layer.
% \begin{figure*}[!t]
% 	\centering
% 	{\includegraphics[width=0.98\linewidth]{rampandattention.pdf}}
% 	\caption{The Architecture of the Hybrid Self-Attention Network}
% 	\label{model}
% \end{figure*}
\subsubsection{Frequency Ramp Structure}
In FEARec, instead of preserving all frequency components, we only extract a subset of frequencies for each layer to guarantee that different attention blocks focus on different spectrums. 
% In order to avoid the frequency missing problem caused by sampling the same frequency range in each layer, we design a dynamic frequency ramp sampling strategy, which means different layers focus on capturing different frequency bands.
% 我们提供两种采样策略：当a>b时，选用重叠采样。当a<b时，
This strategy is used in both time domain attention and frequency domain attention as shown in Figure~\ref{model}.

First, given the input item representation matrix $\mathbf{H}^l \in \mathbb{R}^{N \times D}$ of the $l$-th layer and $\mathbf{H}^{0}=\mathbf{E}_{u}$, we could execute FFT denoted as $\mathcal{F}(\cdot)$ along the item dimension to transform the input item representation matrix $\mathbf{H}^{l}$ to the frequency domain:
\begin{equation}
    \mathcal{F} (\mathbf{H}^{l}) \rightarrow \mathbf{X}^{l}\in \mathbb{C}^{N \times D}
\end{equation}
As said in Appendix~\ref{DFT}, due to the conjugate symmetric property in the frequency domain, only half of the spectrum is used in FEARec
\begin{equation}
    M = {\lceil N / 2\rceil} + 1
\end{equation}
As a result, the sequence length of $\mathbf{X}^{l}$ almost equals half of $\mathbf{H}^{l}$. 
Note that $\mathbf{X}^{l}$ is a complex tensor and represents the spectrum of $\mathbf{H}^{l}$.

\begin{figure}[!t]
	\centering
	{\includegraphics[width=1\linewidth]{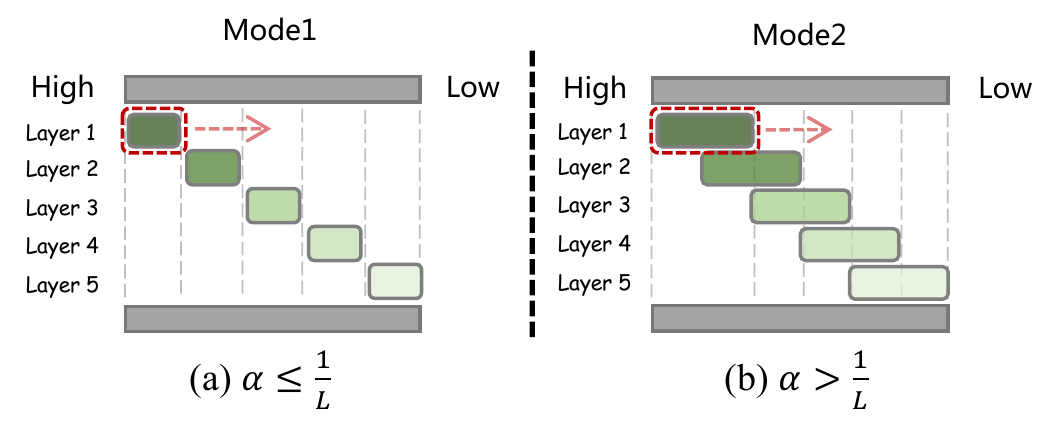}}
	\caption{Frequency ramp sampling. When $\alpha\leq\frac{1}{L}$ sampling mode1 is adopted, and $\alpha >\frac{1}{L}$ sampling mode2 is adopted.}
	\label{Hybrid_att}
\end{figure}

Then, as shown in Figure \ref{Hybrid_att}(a), we gradually select a specific frequency range for each layer and formulate this select operator as
\begin{equation}
    \tilde{\boldsymbol{X}}^l=\operatorname{Sample}_{\alpha}^l(\boldsymbol{X}^l) = \boldsymbol{X}^l[p^l:q^l, :]
\end{equation}
where $\tilde{\boldsymbol{X}}^l \in \mathbb{C}^{F \times D}$ and $F \in \mathbb{N}$ denotes the length of sampled frequency features. We set $\alpha = \frac{F}{M}$ as the initial sampling ratio and all frequency components are retained when $\alpha$ =1. The indexes of the sampled frequency components ($p^l$ and $q^l$) are determined by the position of the current layer in the model. Taking into account that top layers focus more on modeling low-frequency global information while bottom layers are more capable of capturing high-frequency details~\cite{iFormer}, we choose the direction from high frequency to low frequency.
For example, for $l$-th layer: 
\begin{equation}
    p^l=M(1-\alpha)(1-\frac{l-1}{L-1})
\end{equation}
\begin{equation}
    q^l=p^l+\alpha M 
\end{equation}
From Figure \ref{Hybrid_att}(b), when $\alpha > \frac{1}{L}$, the frequencies of different layers are overlapped.
% And when $\alpha \leq \frac{1}{L}$, we adopt average sampling that to ensure all the frequencies are retained:
To avoid the problem of missing frequencies when $\alpha \leq \frac{1}{L}$, we adopt average sampling to ensure that all frequencies are retained:
\begin{equation}
    p^l=M(1-\frac{l}{L})
\end{equation}
\begin{equation}
    q^l= p^l+\frac{M}{L}
\end{equation}
In this way, the whole spectrum is split into several frequency components and preprocessed in different layers. That convenient for us to capture different frequency patterns, and explicitly model high-frequency information that is overlooked by classical self-attention operation. 

\subsubsection{Time Domain Self-Attention Layer}
As shown in Figure \ref{Hybrid_att}, we propose the time domain self-attention to expand the information utilization. For the embedding $\mathbf{H}^{l}$, after linear projector, we get the queries $\boldsymbol{Q}^{l} \in \mathbb{R}^{N \times D}$, keys $\boldsymbol{K}^{l} \in \mathbb{R}^{N \times D}$, and values $\boldsymbol{V}^{l} \in \mathbb{R}^{N \times D}$.
Then we convert $\boldsymbol{Q}^{l}$, $\boldsymbol{K}^{l}$, $\boldsymbol{V}^{l}$ to frequency by FFT and sample a Specific frequency components for each layer. Note that, before computing the attention weights in the time domain, the sampled frequencies need to be zero-padded from $\mathbb{C}^{\alpha M \times D}$ to $\mathbb{C}^{ M \times D}$. The whole process is represented as:
% \begin{equation}
%     \begin{aligned}
%     \tilde{\boldsymbol{Q}}_{s2} &=\operatorname{Sample}_{\alpha_2}(\mathcal{F}(\boldsymbol{Q})) \\
%     \tilde{\boldsymbol{K}}_{s2} &=\operatorname{Sample}_{\alpha_2}(\mathcal{F}(\boldsymbol{K})) \\
%     \end{aligned}
% \end{equation}
\begin{equation}
    \begin{aligned}
    \tilde{\boldsymbol{Q}}^{l}_{t} &=\mathcal{F}^{-1}(\operatorname{Padding}(\operatorname{Sample}^{l}_{\alpha}(\mathcal{F}(\boldsymbol{Q}^{l})))) \\
    \tilde{\boldsymbol{K}}^{l}_{t} &=\mathcal{F}^{-1}(\operatorname{Padding}(\operatorname{Sample}^{l}_{\alpha}(\mathcal{F}(\boldsymbol{K}^{l})))) \\
    \tilde{\boldsymbol{V}}^{l} &=\mathcal{F}^{-1}(\operatorname{Padding}(\operatorname{Sample}^{l}_{\alpha}(\mathcal{F}(\boldsymbol{V}^{l})))) \\
    \end{aligned}
\end{equation}
and
\begin{equation}
    \operatorname{Attention}(\tilde{\boldsymbol{Q}}^{l}_{t}, 
    \tilde{\boldsymbol{K}}^{l}_{t},
     \tilde{\boldsymbol{V}}^{l}) =
     \operatorname{softmax}
     (\frac{\tilde{\boldsymbol{Q}}^{l}_{t}(\tilde{\boldsymbol{K}}^{l}_{t})^{\top}}{\sqrt{D}}) \tilde{\boldsymbol{V}}^{l}
\end{equation}
In this way, our time domain self-attention can learn not only the low-frequency information in the top layers but also the high-frequency in the bottom layers, thus boosting the model's capability of capturing local behaviors.
% We compute the attention with matrix $\tilde{\boldsymbol{Q}}^{l}_{t}$ and $\tilde{\boldsymbol{K}}^{l}_{t}$ transformed back to the time domain, which makes attention focus on a specific frequency range.
% For the multi-head version with the number of heads $h$, multi-head attention concatenate each head as follow.
For the multi-head version used in FEARec, given hidden variables of $D$ channels, $h$ heads, the query, key and value for $i$-th head $\mathcal{Q}_i$, $\mathcal{K}_i$, $\mathcal{V}_i$ $\in \mathbb{R}^{N \times \frac{D}{h}}, i\in\{1,...,h\}$, we have:
\begin{equation}
    \begin{aligned}
    \operatorname{MultiHead}(\mathcal{Q}, \mathcal{K}, \mathcal{V})&=\operatorname{Concat}(\text{head}_1,..., \text{head}_h)\mathcal{W} \\
    \text{where head}_i&= \text {Attention}(\mathcal{Q}_i, \mathcal{K}_i, \mathcal{V}_i)
    \end{aligned}
\end{equation}
Where learnable output matrix $\mathcal{W \in \mathbb{R}^{D \times D}}$. Hence, the multi-head attention operation for $\tilde{\boldsymbol{Q}}^{l}_{t}, 
    \tilde{\boldsymbol{K}}^{l}_{t},
     \tilde{\boldsymbol{V}}^{l}$ in time domain equals to  $\operatorname{MultiHead}_{time}(\tilde{\boldsymbol{Q}}^{l}_{t},
    \tilde{\boldsymbol{K}}^{l}_{t},
     \tilde{\boldsymbol{V}}^{l})$.
% 【关键点】
% \\ \\
% \noindent
% \subsubsection{Frequency Domain Auto-correlation Attention Layer}
\subsubsection{Frequency Domain Attention Layer}
\label{fda}
% As said in the section \ref{introduction}, frequency domain attention discovers the period-based dependencies by calculating the correlation of frequency domain feature and aggregates the most similar behaviors by time delay aggregation.
% Self-attention has been a predominant approach in SR owing to its simplicity and capability of learning sequential dependencies among items.

Existing self-attentive sequential recommenders tend to capture a global view of user interactions at the item level, missing the periodic similar behavior patterns captured at the sequence level.
As discussed in Section~\ref{introduction}, by calculating the auto-correlation, we can find the most related time-delay sequences in the frequency domain and thus discover the periodicity hidden in the behaviors. 
Specifically, as shown as Figure~\ref{introduction}, given a finite sequence of user $s_u = [i_1^{(u)}, i_2^{(u)}, i_3^{(u)}, \cdots, i_{N-1}^{(u)}, i_{N}^{(u)}]$, the time delay operation is defined as $\operatorname{Roll}(s_u,\ \tau)$, where $\tau \in \{1, \cdots, N\}$ indicates the time lag and $\operatorname{Roll}(s_u,\ \tau)$ can be formulated as:
\begin{equation}
    \begin{aligned}
        \operatorname{Roll}(s_u, \tau) &= [i_{\tau + 1}^{(u)}, \cdots, i_{N}^{(u)}, i_{1}^{(u)}, \cdots, i_{\tau}^{(u)}]%  where i_{\tau} mod N
    \end{aligned}
\end{equation}
where
$\operatorname{Roll}(s_u,\ N) = s_u$.
Its corresponding embedding matrix is denoted as $\tilde{\boldsymbol{V}}_{\tau}^{l}$. We perform a similar attention mechanism via auto-correlation in the frequency domain.
For the queries $\boldsymbol{Q}^{l} \in \mathbb{R}^{N \times D}$, keys $\boldsymbol{K}^{l} \in \mathbb{R}^{N \times D}$, and values $\boldsymbol{V}^{l} \in \mathbb{R}^{N \times D}$, 
the auto-correlation is defined as
\begin{equation}
    \begin{aligned}
    \tilde{\boldsymbol{Q}}^{l}_{f} &=\operatorname{Sample}^{l}_{\alpha}(\mathcal{F}(\boldsymbol{Q}^{l})) \\
    \tilde{\boldsymbol{K}}^{l}_{f} &=\operatorname{Sample}^{l}_{\alpha}(\mathcal{F}(\boldsymbol{K}^{l})) \\
    \mathcal{R}_{\tilde{\boldsymbol{Q}}^{l}_{f}, \tilde{\boldsymbol{K}}^{l}_{f}}(\tau)&=\mathcal{F}^{-1}(\operatorname{Padding}(\tilde{\boldsymbol{Q}}^{l}_{f} \odot (\tilde{\boldsymbol{K}}^{l}_{f})^*))
    \end{aligned}
\end{equation}
% Different from the time domain self-attention branch which attend to assign weights to all items in a sequence, frequency domain attention discovers similar sequential patterns from multiple time delay sequences of same user.
% In this section, we first introduce the time delay operation and then calculate the correlation for each delayed sequence.
% Finally, we aggregate the most similar sequential patterns.
% \\ \\
% \noindent
% {\bfseries Time Delay Operation.}
% Different from the time domain self-attention branch which attend to assign weights to all items in a sequence, frequency domain attention attention 
% we perform the  operation to $s_u$ with time delay $\tau$, during which elements that are shifted beyond the first position are reintroduced at the last position as follow:
% Through this way, we change the position of each item and shift the past and recent patterns of user from each time point.
% \\ \\
% \noindent
% {\bfseries Auto-Correlation Attention.}
% After Roll()
% need attention
% 简单聚合位移的序列没有任何意义。
% \mathcal{\boldsymbol{Q},\boldsymbol{K}}}
where $\odot$ represents the element-wise product and * means the conjugate operation. We denote the sampled feature as $\tilde{\boldsymbol{Q}}^{l}_{f}, \tilde{\boldsymbol{K}}^{l}_{f} \in \mathbb{C}^{\alpha M \times D}$. 
$\tilde{\boldsymbol{Q}}^{l}_{f}\odot (\tilde{\boldsymbol{K}}^{l}_{f})^* \in  \mathbb{C}^{\alpha M \times D}$ needs to be zero-padded to $\mathbb{C}^{ M \times D}$ before performing inverse Fourier transform.
% 自相关
Since the information used to compute the attention scores is coming from within the same sequence, here we consider it to be auto-correlation rather than cross-correlation.
The auto-correlation $\mathcal{R}_{\tilde{\boldsymbol{Q}}^{l}_{f}, \tilde{\boldsymbol{K}}^{l}_{f}}(\tau) \in \mathbb{R}^{N \times D}$ is based on the Wiener-Khinchin theorem (more details in the Appendix~\ref{wiener}). 
We further conduct $\operatorname{MEAN}$ operation on $\mathcal{R}_{\tilde{\boldsymbol{Q}}^{l}_{f}, \tilde{\boldsymbol{K}}^{l}_{f}}(\tau)$ to transfer its dimension $\mathbb{R}^{N \times D}$ into $\mathbb{R}^{N}$.

\begin{figure}[!t]
	\centering
	{\includegraphics[width=0.8\linewidth]{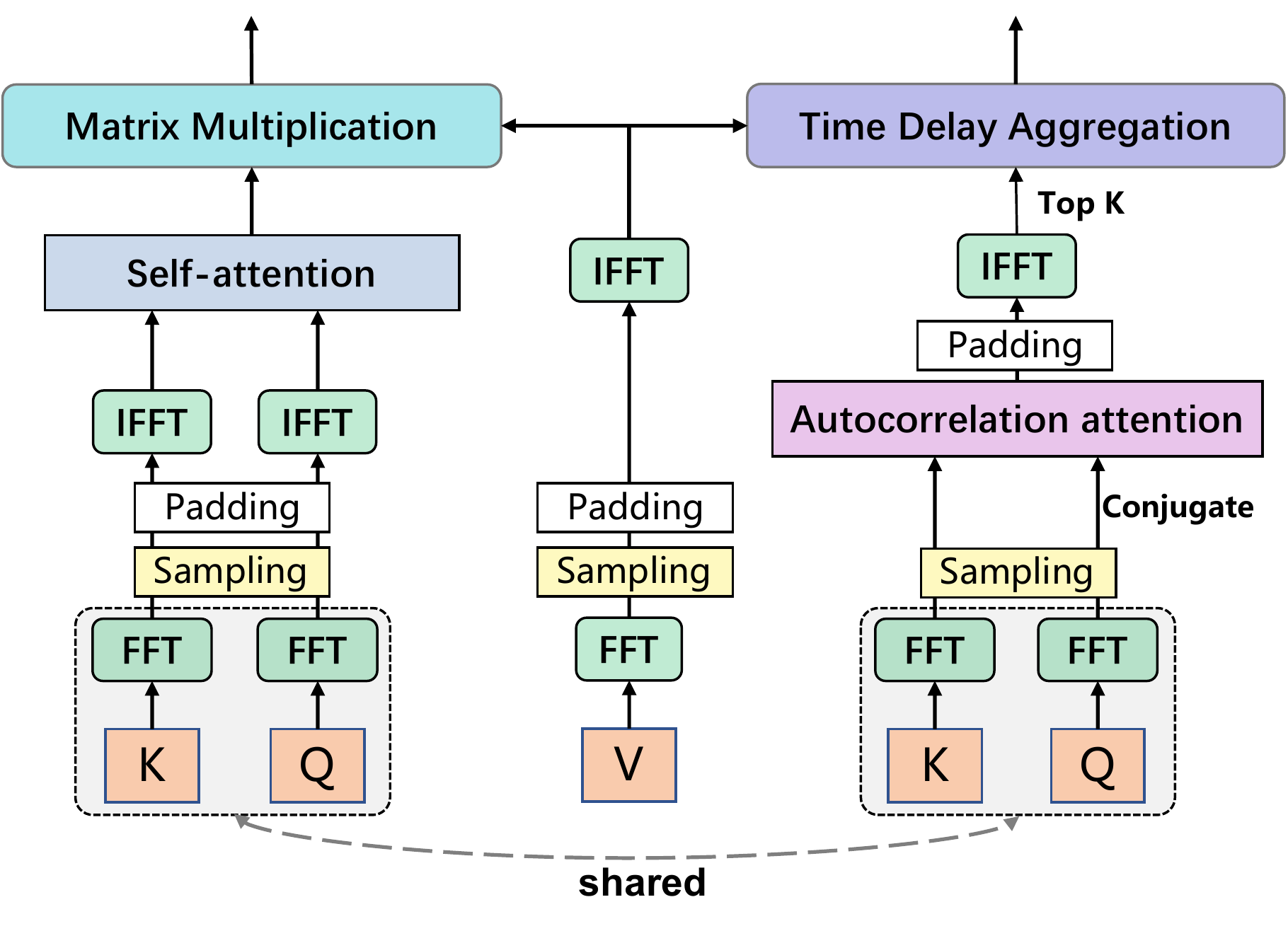}}
	\caption{The architecture of hybrid attention layer.}
	\label{fig:hybrid}
\end{figure}

The auto-correlation refers to the correlation of a time sequence with its own past and future, we choose the most related $\tau_i,...,\tau_k$ from $\mathcal{R}_{\tilde{\boldsymbol{Q}}^{l}_{f}, \tilde{\boldsymbol{K}}^{l}_{f}}(\tau)$:
\begin{equation}
    \tau_1, \cdots, \tau_k =\underset{\tau \in\{1, \cdots, N\}}{\arg \operatorname{Topk}}(\mathcal{R}_{\tilde{\boldsymbol{Q}}^{l}_{f}, \tilde{\boldsymbol{K}}^{l}_{f}}(\tau))
\end{equation}
where $\arg \operatorname{Topk}(\cdot)$ is to get the arguments of the Top $k$ auto-correlation and $k=\lfloor m \times logN\rfloor$.
In this way, we get the most relevant time delay sequences $\operatorname{Roll}(s_u, \tau_i)$, $\tau_i \in \{\tau_1, \cdots, \tau_k\}$. 
Then, the attention weights of different sequences are calculated as:
\begin{equation}
    \begin{aligned}
    \widehat{\mathcal{R}}_{\tilde{\boldsymbol{Q}}^{l}_{f}, \tilde{\boldsymbol{K}}^{l}_{f}}\left(\tau_1\right), \cdots, \widehat{\mathcal{R}}_{\tilde{\boldsymbol{Q}}^{l}_{f}, \tilde{\boldsymbol{K}}^{l}_{f}}\left(\tau_k\right) =\\
    \operatorname{SoftMax}(\mathcal{R}_{\tilde{\boldsymbol{Q}}^{l}_{f}, \tilde{\boldsymbol{K}}^{l}_{f}}\left(\tau_1\right), \cdots, \mathcal{R}_{\tilde{\boldsymbol{Q}}^{l}_{f}, \tilde{\boldsymbol{K}}^{l}_{f}}\left(\tau_k\right))
    \end{aligned}
\end{equation}
Hence, the $k$ most similar sequences can be aggregated by multiplying with their corresponding auto-correlations, which is called \textit{time delay aggregation} in Figure~\ref{fig:hybrid}:
\begin{equation}
    \text { Auto-Correlation }(\tilde{\boldsymbol{Q}}^{l}_{f}, \tilde{\boldsymbol{K}}^{l}_{f},
    \tilde{\boldsymbol{V}}^{l}) =\sum_{i=1}^k \tilde{\boldsymbol{V}}_{\tau_i}^{l} \widehat{\mathcal{R}}_{\tilde{\boldsymbol{Q}}^{l}_{f}, \tilde{\boldsymbol{K}}^{l}_{f}}(\tau_i)
\end{equation}
% $\operatorname{Roll}( \mathcal{X}, \tau)$ represents the operation to $\mathcal{X}$ with time delay $\tau$, during which elements that are shifted beyond the first position are reintroduced at the last position.
where $\text{Auto-Correlation }(\tilde{\boldsymbol{Q}}^{l}_{f}, \tilde{\boldsymbol{K}}^{l}_{f},
\tilde{\boldsymbol{V}}^{l}) \in \mathbb{R}^{N \times D}$. The multi-head attention operation is also conducted for $\tilde{\boldsymbol{Q}}^{l}_{f}, 
    \tilde{\boldsymbol{K}}^{l}_{f},
     \tilde{\boldsymbol{V}}^{l}$ in frequency domain, which equals to  $\operatorname{MultiHead}
     _{Frequency}(\tilde{\boldsymbol{Q}}^{l}_{f}, 
    \tilde{\boldsymbol{K}}^{l}_{f},
     \tilde{\boldsymbol{V}}^{l})$.
% The process is:
% \begin{equation}
%     \begin{aligned}
%     \operatorname{MultiHead}_{Frequency}(\mathcal{Q}, \mathcal{K}, \mathcal{V})&=\operatorname{Concat}(\text{head}_1,..., \text{head}_h)\mathcal{W} \\
% \text{where head}_i&= \text { Auto-Correlation }(\mathcal{Q}_i, \mathcal{K}_i, \mathcal{V}_i).
%     \end{aligned}
% \end{equation}
% Auto-Correlation 是整体的操作
% 【一段有用的话，暂时注释掉】
% Different from the point-wise self-attention family, Auto-Correlation presents the series-wise connections (Figure 3).
% Concretely, for the temporal dependencies, we find the dependencies among sub-series based on the periodicity.
% In contrast, the self-attention family only calculates the relation between scattered points.
% Though some self-attention consider the local information, they only utilize this to help point-wise dependencies discovery. 
% For the information aggregation, we adopt the time delay block to aggregate the similar sub-series from underlying periods.(用过了)
% In contrast, self-attention aggregate the selected points by dot-product.
% Benefiting from the inherent sparsity and sub-series-level representation aggregation, Auto-Correlation can simultaneously benefit the computation efficiency and information utilization.
Finally, we sum the output of the frequency domain attention module and the output of the time domain attention module with the hyperparameter $\gamma$. 
\begin{equation}
    \begin{aligned}
    \widehat{\mathbf{H}}^{l}= &\gamma\operatorname{MultiHead}_{Time}(\tilde{\boldsymbol{Q}}^{l}_{t}, 
    \tilde{\boldsymbol{K}}^{l}_{t},
     \tilde{\boldsymbol{V}}^{l}) +  \\& (1-\gamma)\operatorname{MultiHead}_{Frequency}(\tilde{\boldsymbol{Q}}^{l}_{f}, 
    \tilde{\boldsymbol{K}}^{l}_{f},
     \tilde{\boldsymbol{V}}^{l})
     \end{aligned}
     \label{equal:hybrid}
\end{equation}
% 
% \begin{figure}[!t]
% 	\centering
% 	{\includegraphics[width=0.8\linewidth]{hybrid105.pdf}}
% 	\caption{Adaptive Frequency Ramp Sampling. When $\alpha\leq\frac{1}{L}$ sampling mode1 is adopted, and when $\alpha >\frac{1}{L}$ sampling mode2 is adopted.}
% 	\label{fig:hybrid}
% \end{figure}
% Finally, we sum the output of the frequency domain attention module and the output of the time domain attention module with the hyperparameter $\gamma$. 
% \begin{equation}
%     \begin{aligned}
%     \widehat{\mathbf{H}}^{l}= &\gamma\operatorname{MultiHead}_{Time}(\tilde{\boldsymbol{Q}}^{l}_{t}, 
%     \tilde{\boldsymbol{K}}^{l}_{t},
%      \tilde{\boldsymbol{V}}^{l}) +  \\& (1-\gamma)\operatorname{MultiHead}_{Frequency}(\tilde{\boldsymbol{Q}}^{l}_{f}, 
%     \tilde{\boldsymbol{K}}^{l}_{f},
%      \tilde{\boldsymbol{V}}^{l})
%      \end{aligned}
%      \label{equal:hybrid}
% \end{equation}
% 
\subsubsection{Point-wise Feed Forward Network}
% 10行以内
The frequency domain attention and time domain self-attention are still linear operations, which fail to model complex non-linear relations.
To make the network non-linear, we use a two-layer MLP with a GELU activation function.
The process is defined as follows: 
\begin{equation}
    \tilde{\mathbf{H}}^{l}=\operatorname{FFN}(\widehat{\mathbf{H}}^{l})=(\operatorname{GELU}(\widehat{\mathbf{H}}^{l} \mathbf{W}^{l}_{1}+\mathbf{b}^{l}_{1})) \mathbf{W}^{l}_{2}+\mathbf{b}^{l}_{2}
\end{equation}
where $\mathbf{W}^{l}_{1}, \mathbf{W}^{l}_{2} \in \mathbb{R}^{d\times d}$ and $\mathbf{b}^{l}_{1}, \mathbf{b}^{l}_{2} \in \mathbb{R}^{1\times d}$ are learnable parameters. 
To avoid overfitting, $dropout$ layer, residual connection structure, and layer normalization operations are applied on the obtained output $\mathbf{H}^{l+1}$, as shown below:
\begin{equation}
    \mathbf{H}^{l+1}=\operatorname{LayerNorm}(\mathbf{H}^{l} + \widehat{\mathbf{H}}^{l} + \operatorname{Dropout}(\tilde{\mathbf{H}}^{l}))
\end{equation}
\subsection{Prediction Layer}
% 10行以内
After $L$ FEA blocks that hierarchically extract behavior pattern information of previously interacted items, we get the final combined representation of behavior sequences.
Based on the preference representation $\mathbf{H}^L$, we multiply it by the item embedding matrix $\mathbf{E} \in \mathbb{R}^{|\mathcal{I}| \times D}$ to predict the relevance of the candidate item and use $\operatorname{softmax}$ to convert it into recommendation probability:
\begin{equation}
    \mathbf{\hat{y}}=\operatorname{softmax}(\mathbf{E}^{\top}\mathbf{H}^L)
\end{equation}
Hence, we expect that the true item $i$ adopted by user $u$ can lead to a higher score $\hat{y_i}$.
Therefore, we adopt the cross-entropy loss to optimize the model parameter. The objective function of SR can be formulated as:
\begin{equation}
    \mathcal{L}_{Rec}=-\sum_{i=1}^{|I|} y_{i} \log \left(\hat{y}_{i}\right)+\left(1-y_{i}\right) \log \left(1-\hat{y}_{i}\right)
    \label{LRec}
\end{equation}

% \newpage
% 注释掉
\subsection{Multi-Task Learning}
To enhance the training of hybrid attention to capture 
attention scores and periodic behavior patterns of different frequency bands of user sequence, we leverage a multi-task training strategy to jointly optimize the main recommendation loss with auxiliary dual domain regularization.
% To enhance recommendation with a more supervised signal, we leverage a multi-task training strategy to jointly optimize the main recommendation loss with auxiliary dual domain regularization.
\subsubsection{Contrastive Learning}
Contrastive learning aims to minimize the difference between differently augmented views of the same user and maximize the difference between the augmented sequences derived from different users.
% }
% 【filter】有那么多种对比学习，我们为什么使用模型级对比学习（模型增强）：语义
% Positive Samples

% 无监督对比学习
Although previous augmentations methods~\cite{CL4SRec} including item cropping, masking, and reordering help to enhance the performance of SR models, the data-level augmentations cannot 
% provide a guarantee for high semantic similarity~\cite{DuoRec}.
guarantee a high level of semantic similarity~\cite{DuoRec}.
Instead of using typical data augmentations, we use a dropout-based augmentations methods as shown in the right part of Figure \ref{model}, which is proposed in \cite{R-drop, DuoRec}.
% First when modeling user preferences from input sequences, we let the same user's sequence $s_u$ pass through the network twice.
% \textcolor{blue}{
We let $\mathbf{E}_u$ and $\mathbf{E}_{u}^{\prime}$ pass through the FEA encoder twice for two output views ${\mathbf{H}^{L}_{u}}$ and $({\mathbf{H}^{L}_{u}})^{\prime}$ respectively and model the frequency components to construct harder positive samples by mixing the frequency feature extract from time domain self-attention and frequency autocorrelation attention.
Since there are different dropout layers in the embedding layer and hybrid attention module, we will get two different numerical features but similar semantics.
% }
% 解释为什么还要使用有监督对比学习。
Besides, in order to increase the supervision signal of contrast learning, we follow DuoRec~\cite{DuoRec} and use a sequence $s_{u,s}$ with the same target as $s_u$ as the positive sample of supervised contrast learning.
% Negative Samples (一笔带过！！！)
% To effectively construct the negative samples for an augmented pair of samples, all the other augmented samples in the training batch are considered negative samples (represented as $neg$).
All the other augmented samples in the training batch are treated as negative samples in order to efficiently create the negative samples for an augmented pair of samples (represented as $neg$).

% 【filter】和DuoRec一样，最终我们的loss函数公式如下：
% The contrastive regularization for the batch  $\mathcal{B}$ is defined as: 
For the batch $\mathcal{B}$, the contrastive regularization is defined as:
\begin{equation}
    \mathcal{L}_{\mathrm{CLReg}}=\mathcal{L}_{\mathrm{CLReg}}(\mathbf{h}^{\prime}_u, {\mathbf{h}^{\prime}_{u,s}})+\mathcal{L}_{\mathrm{CLReg}}({\mathbf{h}^{\prime}_{u,s}}, \mathbf{h}^{\prime}_u)
\end{equation}
\begin{equation}
    \mathcal{L}_{\text {CLReg}(\mathbf{h}^{\prime}_u, {\mathbf{h}^{\prime}_{u,s}})}
    =
    -\log \frac{\exp (\operatorname{sim}(\mathbf{h}^{\prime}_u, {\mathbf{h}^{\prime}_{u,s}}))}
    {\sum_{neg} \exp (\operatorname{sim}(\mathbf{h}^{\prime}_u, \mathbf{h}_{neg}))}
    \label{CLReg}
\end{equation}
% 【filter】解释对比学习的公式中的符号和含义
% which computes twice for the unsupervised and the supervised augmented representation respectively. 
% which does two calculations, one for the supervised augmented representation and another for the unsupervised representation.
where $sim(\cdot)$ is dot product and $\mathbf{h}^{\prime}_u$ and $\mathbf{h}^{\prime}_{u,s}$ represent unsupervised and supervised augmented views, respectively, defined as follows:
\begin{equation}
    \mathbf{h}^{\prime}_u=(\mathbf{H}^{L}_u)^{\prime}[-1], \quad {\mathbf{h}^{\prime}_{u,s}}=({\mathbf{H}^{L}_{u,s}})^{\prime}[-1]
\end{equation}

\subsubsection{Frequency Domain Regularization}
% 为了保证增强视图语义的完整性，我们从频域来增强视图。并让它们对齐
The contrastive loss is intuitively performing the push and pull game according to Equation \ref{CLReg} in the time domain.
% 最后来一个频谱dropout就是纯的频率采样。
% alignment of the representations in frequency domain.
Since time-domain and frequency-domain features represent the same semantics, but only in different domains, we assume that the frequency spectrum of similar time-domain features should also be similar.
To ensure the alignment of the representation of different augmented views in the frequency domain,
we suggest an L1 regularization in the frequency domain as a complement to FEARec, which contributes to enriching the regularization of the spectrum of the augmented views.
% \begin{equation}
%     \begin{aligned}
%     \mathcal{L}_{FReg} &=\mathbb{E}[\|\mathcal{F}(\mathbf{h}_u^{\prime})-\mathcal{F}(\mathbf{h}^{\prime}_{u,s})\|_1]
%     \end{aligned}
% \end{equation}

\begin{equation}
    \begin{aligned}
    \mathcal{L}_{FReg} &=\|\mathcal{F}(\mathbf{h}_u^{\prime})-\mathcal{F}(\mathbf{h}^{\prime}_{u,s})\|_1
    \end{aligned}
\end{equation}
% It is worth pointing out that, because of Parseval's relation~\cite{Parseval}, 
% minimizing the MSE in the frequency domain is equivalent to minimizing the MSE in the time domain.
\subsubsection{Train and Inference}
Thus, the overall objective of FEARec with $\lambda$ scale weight is:
\begin{equation}
    \ell = \ell_{\mathrm{Rec}} + \lambda_1\ell_{\mathrm{CReg}} + \lambda_2\ell_{\mathrm{FReg}}
\end{equation}
where $\lambda$ is the hyperparameter to control the strengths of contrastive regularization.
% \textcolor{blue}{
% is the hyperparameter to control the strengths of contrastive regularization.
% is the hyperparameter to control
% }

% \newpage
% 注释掉
\section{EXPERIMENT}
In this section, we first briefly describe the settings in our experiments and then conduct extensive experiments to evaluate our proposed model by answering the following research questions:
\begin{itemize}
    \item {\bfseries RQ1:} How does FEARec perform compared with state-of-the-art SR models?
    \item {\bfseries RQ2:} How do key components, such as frequency sampling, time domain self-attention and frequency domain attention affect the performance of FEARec respectively?
    \item {\bfseries RQ3:} What is the influence of different hyper-parameters in FEARec?
    \item {\bfseries RQ4:} Do frequency domain attention and time domain attention focus on the same features?
\end{itemize}

\subsection{Experimental Setup}
\subsubsection{Dataset}
We conduct experiments on four publicly available benchmark datasets.
\textbf{Beauty, Clothing, and Sports}\footnote{http://jmcauley.ucsd.edu/data/amazon/links.html} are three subsets 
% (\emph{Beauty}, \emph{Clothing Shoes and Jewelry}, and \emph{Sports and Outdoors})
of Amazon Product dataset, which is known for high sparsity and short sequence lengths.
\textbf{MovieLens-1M}\footnote{https://grouplens.org/datasets/movielens/1m/} is a large and dense dataset consisting of long item sequences collected from the movie recommendation site MovieLens.
While \emph{ML-1M} only contains about 1 million interactions.
For all datasets, users/items interacted with less than 5 items/users were removed\cite{Timeinterval, BERT4Rec}.
The statistics of the four datasets after preprocessing are summarized in Table \ref{Datasets}.

\subsubsection{Evaluation Metrics}
In evaluation, we adopt the leave-one-out strategy for each user's item sequence.
As suggested by \cite{KDD}, we rank the predictions over the whole item set without negative sampling.
We report the widely used Top-$n$ metrics \textbf{HR@$n$} (Hit Rate) and \textbf{NDCG@$n$} (Normalized Discounted Cumulative Gain) to evaluate the recommended lists, where $n$ is set to 5, 10.
% In order to ensure the robustness of the experimental results, we repeat each evaluation five times and report the mean value as the final performance.
\begin{table}[t]
\renewcommand{\arraystretch}{0.8}
    \centering
    \caption{Statistics of the datasets after preprocessing.}
    \label{tab:seq}
    \begin{adjustbox}{max width=\linewidth}
        \begin{tabular}{l| r r r r}
        \toprule
        Specs. & Beauty & Clothing & Sports & ML-1M\\
        \midrule
        \midrule
       \# Users & 22,363 & 39,387 & 35,598  &  6,041 \\
       \# Items & 12,101 & 23,033 & 18,357 & 3,417 \\
       \# Avg.Length & 8.9 & 7.1 & 8.3 & 165.5\\
       \# Actions & 198,502 & 278,677 & 296,337 & 999,611\\
       Sparsity & 99.93\% & 99.97\% & 99.95\% & 95.16\%\\
       \bottomrule
    \end{tabular}
    \end{adjustbox}
    \label{Datasets}
\end{table}

\begin{table*}
\renewcommand\arraystretch{1} % 原来用的是1, 0.8太丑
\setlength{\tabcolsep}{0.4em}
    \centering
    \caption{Overall performance over four datasets. Bold scores represent the highest results of all methods. 
    Underlined scores stand for the highest results from previous methods. 
    The FEARec achieves the state-of-the-art result among all baseline models.} %
    \begin{adjustbox}{max width=\textwidth}
        \begin{tabular}{l|l|c| c c c c c |c c c|c|c}

        \toprule
        %\midrule
        Datasets & Metric & BPR-MF & GRU4Rec & Caser & SASRec & BERT4Rec & FMLP-Rec & CL4SRec & CoSeRec & DuoRec & FEARec & Improv.\\
        
        \midrule
        %\midrule
        \multirow{4}{*}{Beauty} 
            &HR@5	&0.0120	&0.0164	&0.0259	&0.0365 &0.0193	&0.0398	&0.0401	&0.0537	&\underline{0.0546}	&\textbf{0.0597}	&9.34\%\\
            &HR@10	&0.0299	&0.0365	&0.0418	&0.0627 &0.0401	&0.0632	&0.0683	&0.0752	&\underline{0.0845}	&\textbf{0.0884}	&4.61\%\\
            &NDCG@5	&0.0040	&0.0086	&0.0127	&0.0236 &0.0187	&0.0258	&0.0223	&0.0361	&\underline{0.0352}	&\textbf{0.0366}	&3.97\%\\
            &NDCG@10	&0.0053	&0.0142	&0.0253	&0.0281 &0.0254	&0.0333	&0.0317	&0.0430	&\underline{0.0443}	&\textbf{0.0459}	&3.61\%\\
            
        % \midrule
        % \multirow{4}{*}{Toys} 
        %     &HR@5	&0.0122	&0.0166	&0.0097	&0.0378	&0.0456	&0.0435	&0	&\underline{0.0670}	&\textbf{0.0701}	&4.63\%\\
        %     &HR@10	&0.0197	&0.0533	&0.0563	&0.0617	&0.0683	&0.0619	&0	&\underline{0.0945}	&\textbf{0.0993}	&5.08\%\\

        %     &NDCG@5	&0.0076	&0.0107	&0.0059	&0.0242	&0.0317	&0.0246	&0	&\underline{0.0393}	&\textbf{0.0430}	&9.41\%\\
        %     &NDCG@10	&0.0100	&0.0224	&0.0359	&0.0319	&0.0391	&0.0305	&0	&\underline{0.0482}	&\textbf{0.0525}	&8.92\%\\
        
        \midrule
        \multirow{4}{*}{Clothing} 
            &HR@5	&0.0067	&0.0095	&0.0108	&0.0168 &0.0125	&0.0173	&0.0168	&0.0175	&\underline{0.0193}	&\textbf{0.0214}	&10.88\%\\
            &HR@10	&0.0094	&0.0165	&0.0174	&0.0272	&0.0208 &0.0277	&0.0266	&0.279	&\underline{0.0302}	&\textbf{0.0323}	&6.95\%\\

            &NDCG@5	&0.0052	&0.0061	&0.0067	&0.0091 &0.0075	&0.0098	&0.0090	&0.0095	&\underline{0.0113}	&\textbf{0.0121}	&7.08\%\\
            &NDCG@10 &0.0069	&0.0083	&0.0098	&0.0124 &0.0102	&0.0127	&0.0121	&0.0131	&\underline{0.0148}	&\textbf{0.0156}	&5.41\%\\
         
        \midrule
        \multirow{4}{*}{Sports}
            &HR@5	&0.0092	&0.0137	&0.0139	&0.0218 &0.0176	&0.0218	&0.0227	&0.0287	&\underline{0.0326}	&\textbf{0.0353}	&8.28\%\\
            &HR@10	&0.0188	&0.0274	&0.0231	&0.0336 &0.0326	&0.0344	&0.0374	&0.0437	&\underline{0.0498}	&\textbf{0.0547}	&9.84\%\\
            &NDCG@5	&0.0040	&0.0096	&0.0085	&0.0127 &0.0105	&0.0144	&0.0129	&0.0196	&\underline{0.0208}	&\textbf{0.0216}	&3.85\%\\
            &NDCG@10	&0.0051	&0.0137	&0.0126	&0.0169 &0.0153	&0.0185	&0.0184	&0.0242	&\underline{0.0262}	&\textbf{0.0272}	&3.82\%\\

        \midrule
        \multirow{4}{*}{ML-1M} 
            &HR@5	&0.0078	&0.0763	&0.0816	&0.1087 &0.0733	&0.1109	&0.1147	&0.1262	&\underline{0.2038}	&\textbf{0.2212}	&8.54\%
            \\
            &HR@10	&0.0162	&0.1658	&0.1593	&0.1904 &0.1323	&0.1932	&0.1975	&0.2212	&\underline{0.2946}	&\textbf{0.3123}	&6.01\%
            \\
            &NDCG@5	&0.0052	&0.0385	&0.0372	&0.0638 &0.0432	&0.0657	&0.0662	&0.0761	&\underline{0.1390}	&\textbf{0.1523}	&9.57\%
            \\
            &NDCG@10	&0.0079	&0.0671	&0.0624	&0.0910 &0.0619	&0.0918	&0.0928	&0.1021	&\underline{0.1680}	&\textbf{0.1861}	&10.77\%\\
            
        % \midrule
        % \multirow{4}{*}{Yelp} 
        %     &HR@5	&0.0127	&0.0152	&0.0156	&0.0161	&0.0179	&0.0216	&0.0241	&\underline{0.0441}	&\textbf{0}	&0\%\\
        %     &HR@10	&0.0245	&0.0263	&0.0252	&0.0265	&0.0304	&0.0352	&0.0395	&\underline{0.0631}	&\textbf{0}	&0\%\\
        %     &NDCG@5	&0.0760	&0.0104	&0.0096	&0.0102	&0.0113	&0.0130	&0.0151	&\underline{0.0325}	&\textbf{0}	&0\%\\
        %     &NDCG@10	&0.0119	&0.0137	&0.0129	&0.0134	&0.0153	&0.0185	&0.0205	&\underline{0.0386}	&\textbf{0}	&0\%\\
        \midrule
        \bottomrule
        \end{tabular}
    \end{adjustbox}
    \label{tab:result}
\end{table*}
\subsubsection{Baselines}
We compare our methods with three types of representative SR models:
\textbf{Non-sequential models:}
BPR-MF~\cite{BPR_MF} is a classic non-sequential method for learning personalized ranking from implicit feedback and optimizing the matrix factorization through a pair-wise Bayesian Personalized Ranking (BPR) loss.
\textbf{Standard sequential models:}
GRU4Rec~\cite{GRU4Rec} is the first recurrent model to apply Gated Recurrent Unit (GRU) to model sequences of user behavior for sequential recommendation.
Caser~\cite{Caser} is a CNN-based method capturing user dynamic patterns by using convolutional filters.
SASRec~\cite{SASRec} 
is a strong Transformer-based model with a multi-head self-attention mechanism.
FMLP-Rec~\cite{FMLP-Rec} is an all-MLP model using a learnable filter-enhanced block to remove noise in the embedding matrix.
\textbf{Sequential models with contrastive learning:}
CL4SRec~\cite{CL4SRec}
generates different contrastive views of the same user interaction sequence for the auxiliary contrastive learning task.
CoSeRec~\cite{CoSeRec} improves the robustness of data augmentation under the contrastive framework by leveraging item-correlations.
DuoRec~\cite{DuoRec} uses unsupervised model-level augmentation and supervised semantic positive samples for contrastive learning. It is the most recent and strongest baseline for the sequential recommendation.
\subsubsection{Implementation Details}
% Baseline
For all baseline models, we reimplement its model and report the result of each model with its optimal hyperparameter settings reported in the original paper.
We implement our FEARec model in PyTorch.
All the experiments are conducted on an NVIDIA V100 GPU with 32GB memory. 
For all datasets, the maximum sequence length $N$ is set to 50.
The dimension of the feed-forward network used in the filer mixer blocks and item embedding size $d$ are both set to 64.
The number of hybrid attention blocks $L$ = 2.
% For the dropout rate on the embedding matrix and hybrid attention blocks are chosen from \{0.1, 0.2, 0.3, 0.4, 0.5\}.
The model is optimized by Adam optimizer with a learning rate of 0.001.
% \textcolor{blue}{
% https://anonymous.4open.science/r/sigirsubmission9566-2164.
% }
% For the adaptive frequency components sampling strategy, we set the $\alpha$ as hyper-parameters and select it from [0, 1] with step 0.1.
% The detailed implementation of our model can be found at https://github.com/FEARec.
\subsection{Overall Performance Comparison (Q1)}
To prove the sequential recommendation performance of our model FEARec, we compare it with other state-of-the-art methods (\textbf{RQ1}).
Table~\ref{tab:result} presents the detailed evaluation results of each model. 

First, it is no doubt that the non-sequential recommendation method BPR-MF displays the lowest results across all datasets since it ignores the sequential information.
Second, compared with the previous RNN-based and CNN-based methods, the advanced Transformer-based method (e.g., SASRec) shows a stronger capability of modeling interaction sequences in SR. 
More recently, a filter-enhanced MLP structure achieved better performance than SASRec by denoising with a learnable filter.
Third, compared with the vanilla method, the model with auxiliary self-supervised learning tasks gains decent improvement. 
The strongest baseline DuoRec outperforms all the previous methods by a large margin by model augmentation and semantic augmentation, which verify the effectiveness of the combination of supervised contrastive learning.
Finally, our model outperforms other competing methods on both sparse and dense datasets with a significant margin across all the metrics, demonstrating the superiority of our model.
% \textcolor{blue}{
In addition, the average improvement on ML-1M is actually larger than that on the Amazon dataset, probably because the average length of users' interaction is extremely short in Amazon.
% 固定指标(指标不算高，不知道要不要单独说)
% Specifically, Model achieves remarkable improvements over the strongest baselines w.r.t. NDCG@5 by 16.83\%, 12.73\%, and 15.20\% in Amazon Clothing, ML-1M, and Yelp, respectively. 
% 结论要改：
This observation demonstrates the effectiveness of FEARec in both sparse and dense dataset and shows that transforming user sequence from the time domain to the frequency domain and improving the original attention mechanism with frequency ramp structure and extra frequency domain attention is promising to extract useful information for accurate recommendation.
% }
% 新的消融实验表格
% begin{table}[!t]
\begin{table}[!t]
\renewcommand{\arraystretch}{1.1}
    \centering
    \caption{Ablation study of FEARec in terms of HR@5 and NDCG@5 on Beauty, Clothing and ML-1M datasets.}
    \label{tab:seq5}
    \begin{adjustbox}{max width=\linewidth}
        \begin{tabular}{lccccccccc}
            \toprule
            \multicolumn{1}{c}{\multirow{2}{*}{Methods}} &
            &
            % \multicolumn{1}{c}{} &
            \multicolumn{2}{c}{Beauty} & 
            & 
            \multicolumn{2}{c}{Clothing} & 
            & 
            \multicolumn{2}{c}{ML-1M} \\ 
            \cmidrule{3-4}
            \cmidrule{6-7}
            \cmidrule{9-10}
            % \cmidrule{2-7}
            % \hline
            & & HR@5 & NDCG@5 & & HR@5 & NDCG@5 & & HR@5 & NDCG@5 \\
            \hline
            \hline
            FEARec & &\textbf{0.0597}&\textbf{0.0366} & &\textbf{0.0208} &\textbf{0.0119} & &\textbf{0.2212} &\textbf{0.1523} \\
            \hline
            (a) w/o STD & &0.0578 &0.0356 & &0.0214 &0.0117 & &0.2141 &0.1482 \\
            (b) w/o SFD & &0.0566 &0.0348 & &0.0204 &0.0115 & &0.2134 &0.1457 \\
            \hline
            (c) w/o TDA & &0.0589 &0.0358 & &0.0203 &0.0113 & &0.2123 &0.1441 \\
            (d) w/o FDA & &0.0590 &0.0361 & &0.0202 &0.0116 & &0.2116 &0.1470 \\
            \hline
            (e) w/o FReg & &0.0582 &0.0360 & &0.0208 &0.0119 & &0.2114 &0.1433 \\
            (f) w/o CReg & &0.0567 &0.0342 & &0.0191 &0.0106 & &0.2028 &0.1408 \\
            \bottomrule
        \end{tabular}
    \end{adjustbox}
    \label{table:ablation}
\end{table}
\subsection{Ablation Study (Q2)}
In this section, we conduct ablation studies to evaluate the effectiveness of each key component.
% including sampling in time domain attention module (STD), sampling in frequency domain attention module (SFD), time domain attention (TDA), frequency domain attention (FDA), frequency domain regularization (FReg) and contrastive regularization (CReg).
Table \ref{table:ablation} shows the performance of our default method and its 6 variants on three datasets.
% We introduce the variants and analyze their effect respectively.
% 【1013】
% In this section, we conduct ablation studies to evaluate the effectiveness of each key component, including frequency ramp sampling, hybrid attention layer, contrastive loss and frequency domain regularization.
% Table \ref{table:ablation} shows the performance of our default method and its 5 variants on three datasets.
% We introduce the variants and analyze their effect respectively.
% 【1013】
% \subsubsection{Frequency components sampling}
% In hybrid attention block, the frequency component of the input feature needs to be selected via frequency ramp structure, and then input to the time-domain and frequency-domain attention modules respectively to calculate attention in item-level and sub-series level.

\textbf{Frequency Ramp Structure.}
% 实验设置
In order to verify the effectiveness of the frequency ramp structure sampling in FEARec, we remove the sampling structure from the time-domain attention module (w/o STD) and frequency-domain (w/o SFD), which means that sets the sampling ratio $\alpha$ equals 1, that is, all frequency components are retained.
% Compared with (a)-(b), we find that without frequency ramp sampling in time domain or frequency domain, the performance drops significantly, which verifies our claim that preserving all frequency components will result in suboptimal results and the structure can trade-off high-frequency and low-frequency components across all layers.
% 实验现象：
% \textcolor{blue}{
Compared with (a)-(b), we find that without frequency ramp sampling in time domain or frequency domain, the performance drops significantly, which verifies the structure can trade-off high-frequency and low-frequency components across all layers.
% Finally, capture local high-frequency information at botton layer then gradually get a global view of user’s interaction data and the overall evolution of user preference. 
In summary, high-frequency information is captured at the bottom layer, then low-frequency information is gradually captured, and finally, the overall evolution of user preferences is obtained.
% }

% and both time and frequency domain attention learn better with the sampling strategy.

% \subsubsection{Hybrid dual domain attention}

\textbf{Hybrid Attention Module.}
From (c)-(d) we can see that each attention module plays a crucial role in modeling user preference, and removing time domain attention
(w/o TDA) or frequency domain attention (w/o FDA) leads to a performance decrease.
% Besides, we can see that FEARec outperforms all the other variants with a single attention module.
% \textcolor{blue}{
The time-domain self-attention layer captures the attention score in the item level.
However, the frequency domain attention layer computes the auto-correlation  to recognize the periodic characterize in the sub-sequence level.
The hybrid attention module combines two attention modules into a unified structure as illustrated in Figure~\ref{fig:hybrid}, which is promising to better capture user preference.
% 应该是快要到右半边的第二行！！！
% }
\begin{figure*}[t]
	\centering
	{\includegraphics[width=1\linewidth]{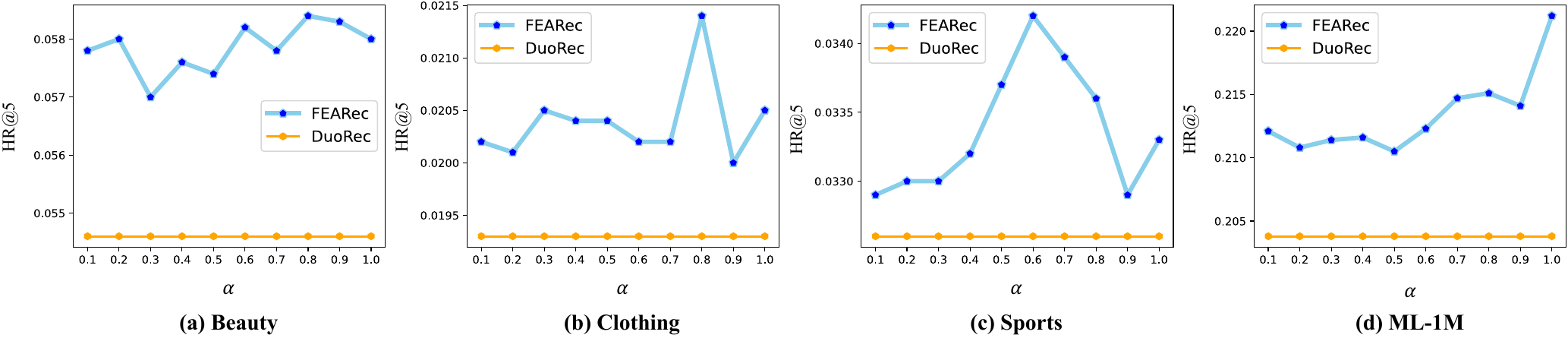}}
	\caption{Performance (HR@5) comparison w.r.t different sampling ratio on four datasets.}
	\label{sample_ratio}
\end{figure*}
\begin{figure}[t]
	\centering
	{\includegraphics[width=1\linewidth]{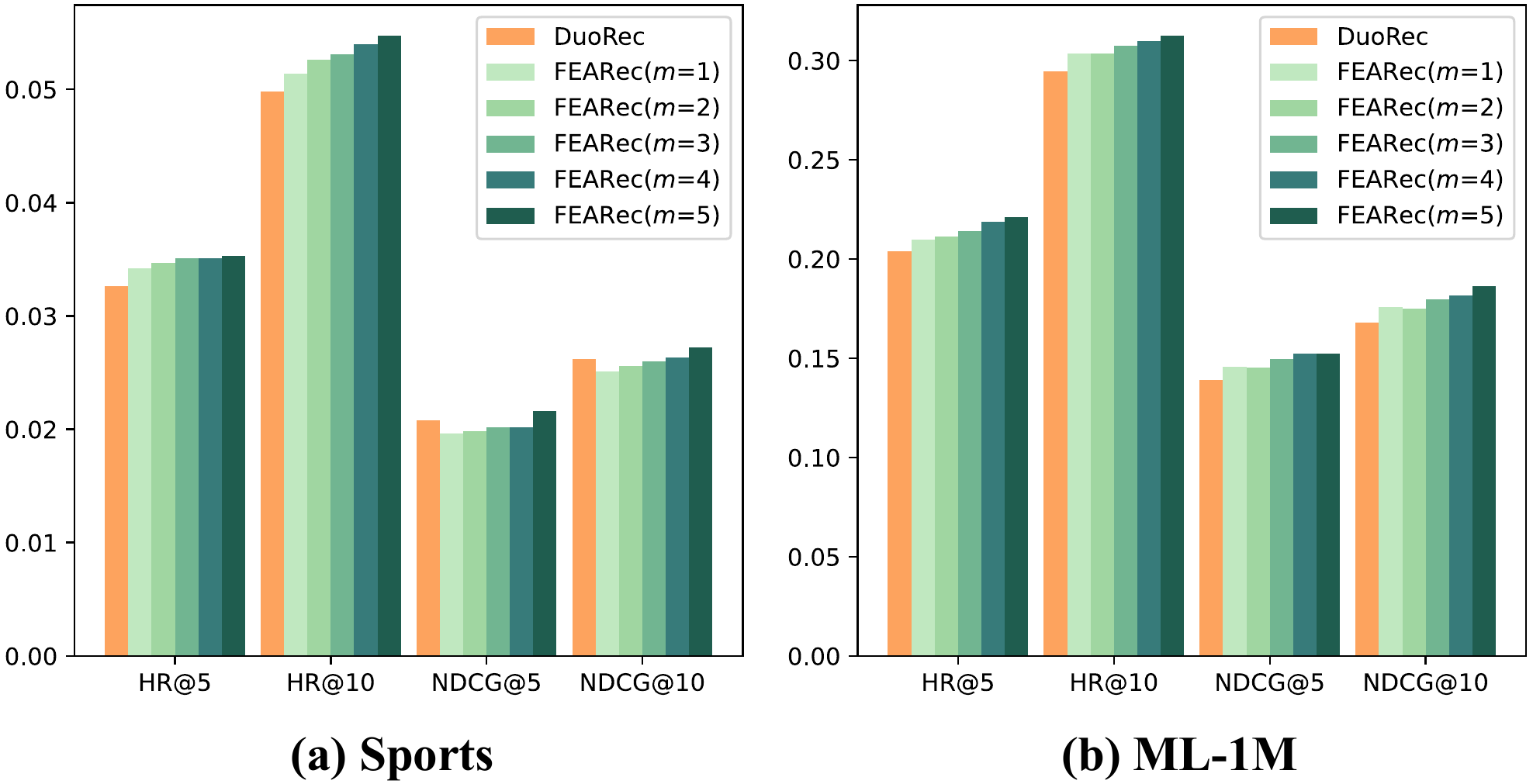}}
	\caption{Performance comparison w.r.t different top $k$ time delay aggregation on Sports and ML-1M datasets.}
	\label{fig:topk}
\end{figure}
% \subsubsection{Contrastive Learning and regularization}

\textbf{Contrastive Learning and regularization.}
From (e)-(f) we can see that combining contrastive learning regularization and a simple frequency domain regularization will improve the performance.
When removing the contrastive auxiliary task of contrastive learning will hurt performance, which is consistent with the previous observation that sequential recommendation benefits from contrastive learning \cite{CL4SRec,CoSeRec, DuoRec}.
As a complement to contrastive learning, frequency domain regularization can implicitly reduce the distance between the spectrum of two augmented perspectives, but adopting it alone will yield poor results.
%     Interestingly, on ML-1M (with the longest average user sequences), model-based encoders exceed all (2) Presumably because RNN encodes actions 'one-by-one' and cannot capture more felxible 'skip' behaviors like other encoders, RNN performs the worst on these three datasets.
% \end{itemize}

\begin{table}[t]
\renewcommand{\arraystretch}{0.8}
    \centering
    \caption{Performance (HR@5) of time-domain self-attention (TDA) and frequency-domain attention (FDA) module mixed in different proportions on four datasets.}
    \label{tab:hybrid_ratio}
    \begin{adjustbox}{max width=\linewidth}
        \begin{tabular}{c|c c| c c c c}
        \toprule
        Hybrid & TDA & FDA & Beauty  & Clothing & Sports & ML-1M\\
        \midrule
        \midrule
        %\multirow{4}{*}{Slide Mode} 
        (1)
        &0.1 &0.9 &\textbf{0.0596} &\textbf{0.0214} &\textbf{0.0349} &0.1863\\
        
        (2)
        &0.3 &0.7 &0.0584 &0.0204 &0.0335 &0.2113\\
        
        (3)
        &0.5 &0.5 &0.0557 &0.0207 &0.0346 &0.2154\\
        
        (4)
        &0.7 &0.3 &0.0591 &0.0203 &0.0342 &0.2169\\
        
        (5)
        &0.9 &0.1 &0.0592 &0.0211 &0.0339 &\textbf{0.2212}\\
       \bottomrule
    \end{tabular}
    \end{adjustbox}
\end{table}

\subsection{Hyper-parameter Sensitivity (Q3)}
In this section, we study the influence of three important hyper-parameters in FEARec, including sampling ratio $\alpha$, hybrid ratio $\gamma$ and autocorrelation top $k$. 
We keep all other hyperparameters optimal when investigating the current hyperparameter.
% To control variables, we only change one hyper-parameter at one time while keeping others optimal.

% \subsubsection{Module Hybrid Ratio $\gamma$}
\textbf{Module Hybrid Ratio $\gamma$.}
In order to simultaneously model user preferences at the item level and sub-sequence level, we mix the original TDA and FDA through Figure \ref{fig:hybrid}, and the weight of the frequency domain and time domain is controlled by Eq.~(\ref{equal:hybrid}).
% 分析
From the table \ref{tab:hybrid_ratio}, we can see that on the Beauty, Clothing and Sports dataset, the best effect is achieved when the FDA module is assigned greater weight than the TDA module, which verifies the effectiveness of frequency domain attention.
While on the dense ML-1M dataset, the TDA module plays a more important role.

\textbf{Sampling Ratio $\alpha$.}
% One of the novelties in the proposed FEARec model is that the frequency component of the input feature needs to be selected via frequency ramp structure and then input to the time-domain and frequency-domain attention modules respectively to calculate attention in item-level and sub-series level.
% 介绍这个超参数：
% \textcolor{blue}{
The ratio of frequency components is controlled by $\alpha$, which significantly affects how much of the frequency component is retained.
Although the average length of user interactions varies across data sets, the number of all frequency components in the frequency domain is $M$.
% }
We conduct experiments under different $\alpha$ on four datasets.
% and illustrate the impact of this hyperparameter in Figure \ref{sample_ratio}. 
As shown in Figure \ref{sample_ratio}, we observe that with the increase of $\alpha$ the performance of FEARec starts to increase at the beginning, and it gradually reaches its peak when $\alpha$ equals 0.8 on Amazon-Beauty and Amazon-Clothing, and 0.6 on Amazon-Sports.
Afterward, it starts to decline.
While on dense datasets like ML-1M, to capture more complex sequential patterns of users, the hybrid attention structure requires more frequency components.
% in the frequency domain.
% No matter what the sampling values of $\alpha$, our method FEARec always performs better than DuoRec.
Our method FEARec always performs better than DuoRec, regardless of the value of $\alpha$.

% \textbf{Sampling Ratio $\alpha$.}
% % One of the novelties in the proposed FEARec model is that the frequency component of the input feature needs to be selected via frequency ramp structure and then input to the time-domain and frequency-domain attention modules respectively to calculate attention in item-level and sub-series level.
% % 介绍这个超参数：
% \textcolor{blue}{
% The ratio of frequency components is controlled by $\alpha$, which significantly affects how much of the frequency component is retained.
% Although the average length of user interactions varies across data sets, the number of all frequency components in the frequency domain is $M$.
% }
% We conduct experiments under different $\alpha$ on four datasets and illustrate the impact of this hyperparameter in Figure \ref{sample_ratio}. 
% As shown in Figure \ref{sample_ratio}, we observe that with the increase of $\alpha$ the performance of FEARec starts to increase at the beginning, and it gradually reaches its peak when $\alpha$ equals 0.8 on Amazon-Beauty and Amazon-Clothing, and 0.6 on Amazon-Sports.
% Afterward, it starts to decline.
% While on dense datasets like ML-1M, to capture more complex sequential patterns of users, the hybrid attention structure requires more frequency components in the frequency domain.
% Note that for all the sampling values of $\alpha$, our method FEARec always performs better than DuoRec.

% 根据最终图的效果来确定：
% \subsubsection{Auto-correlation Top$k$}
\textbf{Auto-correlation Top-$k$.}
In frequency domain attention, we need to aggregate the Top-$k$ most relevant time delay sequence, where $k$ is determined by the parameter $m$ in section~\ref{fda}.
% In order to verify the effectiveness of frequency domain attention, we gradually increase the number of time delayed sequences that are most similar to the original sequence. 
% From Figure \ref{fig:topk}, we can find that with the increase of $f$, more sequences with higher autocorrelation are aggregated together by the frequency domain attention module, which can effectively improve the performance of FEARec.
We gradually increase the parameter $m$ to aggregate more time delayed sequences with high autocorrelation. As shown in Figure \ref{fig:topk}, we can observe that the performance of FEARec consistently improves as we increase $m$ up to a certain point. However, it is worth noting that further increasing $m$ beyond a certain value will result in performance degradation, which is not shown in the figure. This is because when $m$ is too large, the frequency domain attention module may start to include irrelevant time delayed sequences, which can negatively impact the model's performance. 
% Therefore, it is important to carefully choose the value of $f$ and use the top-$k$ filtering mechanism to ensure that only the most relevant time delayed sequences are considered by the frequency domain attention module. Overall, our experiments demonstrate that FEARec's frequency domain attention mechanism is effective in capturing periodic characteristics, and the top-$k$ filtering mechanism is necessary to prevent the inclusion of irrelevant time delayed sequences.

% \textcolor{blue}{
% which can effectively improve the performance of FEARec.
% which can effectively improve the performance of FEARec.
% which can effectively improve the performance of FEARec.
% }

\begin{figure}[t]
	\centering
	{\includegraphics[width=1\linewidth]{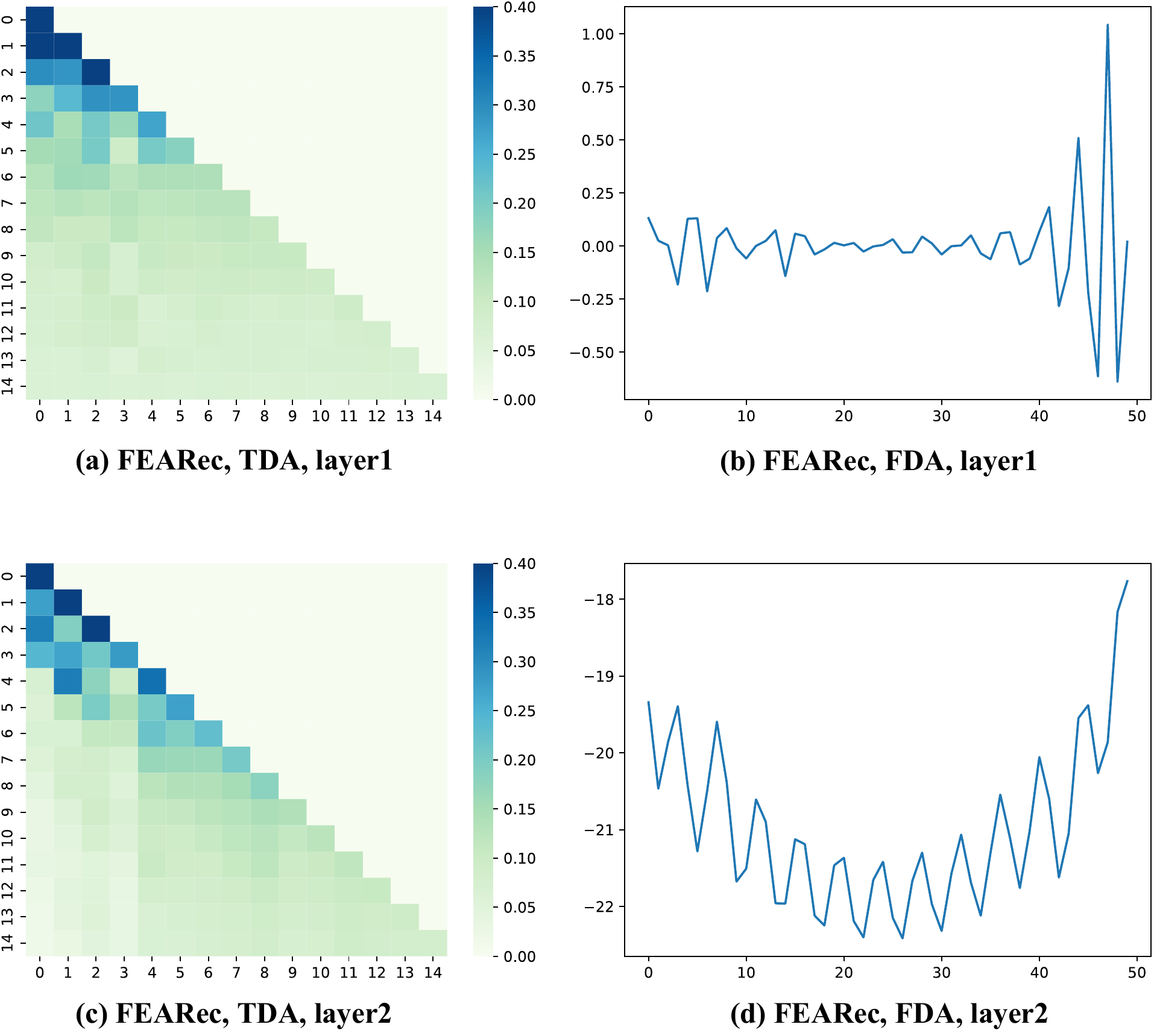}}
	\caption{Average attention weights and autocorrelation scores of FEARec at different layers on the ML-1M dataset.}
	\label{fig:visualization}
\end{figure}
\subsection{Attention Visualization (Q4)}
To evaluate how the hybrid attention capture sequential behavior in both item-level and sub-sequence level, the attention score and autocorrelation score learned in TDA and FAD in each layer are visualized on the ML-1M dataset.
The results are displayed in Figure \ref{fig:visualization}.
Based on our observation, we can derive the following results: 
% the time-domain attention in the hybrid attention module tends to assign more weight to the items initially interacted, because the user has established his basic preference among the multiple items initially interacted, and these items constitute the most critical interaction mode.
% 这句话意思可能不对，attention score是item之间的关系。
the time-domain attention in the hybrid attention module tends to assign more weight to the items initially interacted with because the users established their basic preference among these items, which constitute a critical sequential pattern of the user.
Different from the time-domain attention of item level, the visualization results of frequency-domain attention represent the autocorrelation weights of different time delay sequences.
% \textcolor{blue}{
It can be seen from Figure \ref{fig:topk}(b) and (d) that if the time delay is small or large for the original sequence, a higher autocorrelation score will be obtained, which effectively show periodic characteristics on ML-1M datasets.
% }
% From (b) we observe that the feature of sequence delayed by $\tau$=48 is perfectly correlated to the original sequence that with no delay.
% In that case the original feature are shown to be periodic.

% \newpage
% 注释掉
\section{CONCLUSION}
% The widely used self-attention operation is limited by its capability in modeling high-frequency information for sequential recommendation.
% Meanwhile, the inherent periodicity is hard for these time domain based methods to recognize.
% Hence, it is intuitive to alleviate the above problems in the frequency domain.
In this paper, we design a new frequency-based model, FEARec, to build a hybrid attention manner in both the time and frequency domains.
By working upon our defined frequency ramp structure, FEARec employs improved time domain attention to learn both low-high frequency information.
We also propose a frequency domain attention to exploring inherent dependencies existing in the sequential behaviors by calculating the autocorrelation of different time delay sequences.
We further adapt contrastive learning and frequency domain regularization to ensure that multiple views are aligned. 
Lastly, due to the capability of modeling different frequency information and periodic characteristics, FEARec show the superiority over all state-of-the-art models.

\begin{acks}
% This research was partially supported by the NSFC(61876117, 618762
% 17, 62176175), the major project of natural science research in Universities of Jiangsu Province(21KJA520004) and Project Funded by the Priority Academic Program Development of Jiangsu Higher Education Institutions.
This research was partially supported by the NSFC (61876117, 62176175), the major project of natural science research in Universities of Jiangsu Province (21KJA520004), Suzhou Science and Technology Development Program(SYC2022139), the Priority Academic Program Development of Jiangsu Higher Education Institutions.
\end{acks}

% \newpage
% \bibliographystyle{IEEEtran}
% \bibliography{references}{}

% \newpage
\appendix
\section{Appendix}
\subsection{Fourier Transform}
\subsubsection{Discrete Fourier Transform}
\label{DFT}
% The Fourier transform is a bridge between the time domain and the frequency domain of the signal. 
Discrete Fourier Transform (DFT) is one of the most widely used computing methods, with numerous applications in data analysis, signal processing, and machine learning~\cite{DFT1, DFT2}.
As the input data of SR is one-dimensional sequences, only the 1D DFT is considered in our FEARec. Given a finite sequence $\{x_n\}_{n=1}^{N}$, the 1D DFT could convert the original sequence into the sequence of complex numbers in the frequency domain by:
\begin{equation}
    \label{eq1}
    X_k=\sum_{n=1}^{N} x_n W_N^{nk}, \quad 1 \leq k \leq N
\end{equation}
where $N$ is the length of the sequence, $W_N^{nk}$ is the twiddle factor, and $X_k$ is a complex number that represents the signal with frequency $\omega_k=2\pi k/N$.
Through Eq.~(\ref{eq1}), the DFT is completed by decomposing a sequence of values into components of different frequencies. Note that DFT is a one-to-one unique mapping operation in the time and frequency domains. And the sequence of frequency representation $\{X_k\}_{k=1}^{N}$ can be transferred to the original feature domain via an inverse DFT (IDFT), which is formulated as:
\begin{equation}
    x_{n}=\frac{1}{N} \sum_{k=1}^{N} X_{k} W_N^{-nk}
\end{equation}
For real input $x_n$, it has been proven that its DFT is conjugate symmetric, i.e., $X_k=X^*_{N-k}$, where $*$ denotes the conjugate operation. 
It indicates that the half of the DFT $ \{X_k\}_{k=0}^{\lceil N / 2\rceil}$ contains the full information about the frequency characteristics of $x_n$. 
If we perform IDFT to $ \{X_k\}_{k=0}^{\lceil N / 2\rceil}$, a real discrete signal can be recovered. 

\subsubsection{Fast Fourier Transform}
\label{FFT}
The Fast Fourier Transform (FFT) algorithm~\cite{cooley,FFTW3} is a fast algorithm for computing the DFT of a sequence, which takes advantage of the symmetry and periodicity properties of $W_N^{kn}$ and reduces the complexity to compute DFT from $\mathcal{O}(N^2)$ to $\mathcal{O}(N\log N)$. 
The Inverse FFT (IFFT), which has a similar form to the DFT, can also be used to efficiently compute the time features corresponding to $x_{n}$. 
In this paper, we denote FFT and IFFT by  $\mathcal{F}$ and $\mathcal{F}^{-1}$, respectively.

% \subsection{Autocorrelation}
\subsection{Wiener-Khinchin Theorem}
\label{wiener}
Given a discrete-time sequence $\mathcal{X}_n=\{x_n\}_{n=1}^{N}$, we can obtain the auto-correlation $\mathcal{R}_{\mathcal{X X}}(\tau)$ in the time domain by the following equations:
\begin{equation}
    \mathcal{R}_{\mathcal{X} \mathcal{X}}(\tau)=\lim _{N \rightarrow \infty} \frac{1}{N} \sum_{n=1}^{N} \mathcal{X}_n \mathcal{X}_{n-\tau}
\end{equation}
$\mathcal{R}_{\mathcal{X X}}(\tau)$ reflects the time-delay similarity between $\left\{\mathcal{X}_n\right\}$ and its $\tau$ lag series $\left\{\mathcal{X}_{n-\tau}\right\}$.
% Therefore
Auto-correlation is an ideal method for uncovering trends and patterns in time series data.
Such merit is particularly appealing for the recommendation task, where
users’ behaviors tend to show certain periodic trends \cite{merit_11, merit_19, merit_21}. 
% Q, K, V

% 我们在频率进行计算。
To reduce the complexity of auto-correlation computation, the $\mathcal{R}_{\mathcal{X X}}(\tau)$ is calculated in the frequency domain by FFT based on the Wiener-Khinchin theorem~\cite{wiener}.
\begin{equation}
\begin{aligned}
&\mathcal{S}_{\mathcal{X X}}(f)=\mathcal{F}\left(\mathcal{X}_n\right) \mathcal{F}^*\left(\mathcal{X}_n\right) \\
&\mathcal{R}_{\mathcal{X X}}(\tau)=\mathcal{F}^{-1}\left(\mathcal{S}_{\mathcal{X X}}(f)\right)
\end{aligned}
\end{equation}
where $\tau \in \{1,...,N\}$, $\mathcal{F}$ denotes the FFT and $\mathcal{F}^{-1}$ is its inverse.
And $\mathcal{S}_{\mathcal{X} \mathcal{X}}(f)$ is in the frequency domain.
Note that the series auto-correlation of all lags in $\{1,...,N\}$ can be calculated at once by FFT.
Thus, auto-Correlation achieves the $\mathcal{O}(N\log N)$ complexity.

% \newpage
\bibliographystyle{IEEEtran}
\bibliography{references}{}

% Generated by IEEEtran.bst, version: 1.14 (2015/08/26)
\begin{thebibliography}{10}
\providecommand{\url}[1]{#1}
\csname url@samestyle\endcsname
\providecommand{\newblock}{\relax}
\providecommand{\bibinfo}[2]{#2}
\providecommand{\BIBentrySTDinterwordspacing}{\spaceskip=0pt\relax}
\providecommand{\BIBentryALTinterwordstretchfactor}{4}
\providecommand{\BIBentryALTinterwordspacing}{\spaceskip=\fontdimen2\font plus
\BIBentryALTinterwordstretchfactor\fontdimen3\font minus
  \fontdimen4\font\relax}
\providecommand{\BIBforeignlanguage}[2]{{%
\expandafter\ifx\csname l@#1\endcsname\relax
\typeout{** WARNING: IEEEtran.bst: No hyphenation pattern has been}%
\typeout{** loaded for the language `#1'. Using the pattern for}%
\typeout{** the default language instead.}%
\else
\language=\csname l@#1\endcsname
\fi
#2}}
\providecommand{\BIBdecl}{\relax}
\BIBdecl

\bibitem{BERT4Rec}
F.~Sun, J.~Liu, J.~Wu, C.~Pei, X.~Lin, W.~Ou, and P.~Jiang, ``Bert4rec:
  Sequential recommendation with bidirectional encoder representations from
  transformer,'' in \emph{Proceedings of the 28th ACM international conference
  on information and knowledge management}, 2019, pp. 1441--1450.

\bibitem{GRU4Rec}
D.~Jannach and M.~Ludewig, ``When recurrent neural networks meet the
  neighborhood for session-based recommendation,'' in \emph{Proceedings of the
  eleventh ACM conference on recommender systems}, 2017, pp. 306--310.

\bibitem{Markov}
R.~He and J.~McAuley, ``Fusing similarity models with markov chains for sparse
  sequential recommendation,'' in \emph{2016 IEEE 16th international conference
  on data mining (ICDM)}.\hskip 1em plus 0.5em minus 0.4em\relax IEEE, 2016,
  pp. 191--200.

\bibitem{Attention}
A.~Vaswani, N.~Shazeer, N.~Parmar, J.~Uszkoreit, L.~Jones, A.~N. Gomez,
  {\L}.~Kaiser, and I.~Polosukhin, ``Attention is all you need,''
  \emph{Advances in neural information processing systems}, vol.~30, 2017.

\bibitem{SASRec}
W.-C. Kang and J.~McAuley, ``Self-attentive sequential recommendation,'' in
  \emph{2018 IEEE international conference on data mining (ICDM)}.\hskip 1em
  plus 0.5em minus 0.4em\relax IEEE, 2018, pp. 197--206.

\bibitem{CoSeRec}
Z.~Liu, Y.~Chen, J.~Li, P.~S. Yu, J.~McAuley, and C.~Xiong, ``Contrastive
  self-supervised sequential recommendation with robust augmentation,''
  \emph{arXiv preprint arXiv:2108.06479}, 2021.

\bibitem{DuoRec}
R.~Qiu, Z.~Huang, H.~Yin, and Z.~Wang, ``Contrastive learning for
  representation degeneration problem in sequential recommendation,'' in
  \emph{Proceedings of the Fifteenth ACM International Conference on Web Search
  and Data Mining}, 2022, pp. 813--823.

\bibitem{CL4SRec}
X.~Xie, F.~Sun, Z.~Liu, S.~Wu, J.~Gao, J.~Zhang, B.~Ding, and B.~Cui,
  ``Contrastive learning for sequential recommendation,'' in \emph{2022 IEEE
  38th International Conference on Data Engineering (ICDE)}.\hskip 1em plus
  0.5em minus 0.4em\relax IEEE, 2022, pp. 1259--1273.

\bibitem{lspp}
C.~Xu, J.~Feng, P.~Zhao, F.~Zhuang, D.~Wang, Y.~Liu, and V.~S. Sheng,
  ``Long-and short-term self-attention network for sequential recommendation,''
  \emph{Neurocomputing}, vol. 423, pp. 580--589, 2021.

\bibitem{FDSA}
T.~Zhang, P.~Zhao, Y.~Liu, V.~S. Sheng, J.~Xu, D.~Wang, G.~Liu, and X.~Zhou,
  ``Feature-level deeper self-attention network for sequential
  recommendation.'' in \emph{IJCAI}, 2019, pp. 4320--4326.

\bibitem{MMInfoRec}
R.~Qiu, Z.~Huang, and H.~Yin, ``Memory augmented multi-instance contrastive
  predictive coding for sequential recommendation,'' in \emph{2021 IEEE
  International Conference on Data Mining (ICDM)}.\hskip 1em plus 0.5em minus
  0.4em\relax IEEE, 2021, pp. 519--528.

\bibitem{ICLRec}
Y.~Chen, Z.~Liu, J.~Li, J.~McAuley, and C.~Xiong, ``Intent contrastive learning
  for sequential recommendation,'' in \emph{Proceedings of the ACM Web
  Conference 2022}, 2022, pp. 2172--2182.

\bibitem{howVITwork}
N.~Park and S.~Kim, ``How do vision transformers work?'' \emph{arXiv preprint
  arXiv:2202.06709}, 2022.

\bibitem{VIT_seelike_CNN}
M.~Raghu, T.~Unterthiner, S.~Kornblith, C.~Zhang, and A.~Dosovitskiy, ``Do
  vision transformers see like convolutional neural networks?'' \emph{Advances
  in Neural Information Processing Systems}, vol.~34, pp. 12\,116--12\,128,
  2021.

\bibitem{iFormer}
C.~Si, W.~Yu, P.~Zhou, Y.~Zhou, X.~Wang, and S.~Yan, ``Inception transformer,''
  \emph{arXiv preprint arXiv:2205.12956}, 2022.

\bibitem{LOCKER}
Z.~He, H.~Zhao, Z.~Lin, Z.~Wang, A.~Kale, and J.~McAuley, ``Locker: Locally
  constrained self-attentive sequential recommendation,'' in \emph{Proceedings
  of the 30th ACM International Conference on Information \& Knowledge
  Management}, 2021, pp. 3088--3092.

\bibitem{LSAN}
Y.~Li, T.~Chen, P.-F. Zhang, and H.~Yin, ``Lightweight self-attentive
  sequential recommendation,'' in \emph{Proceedings of the 30th ACM
  International Conference on Information \& Knowledge Management}, 2021, pp.
  967--977.

\bibitem{FMLP-Rec}
K.~Zhou, H.~Yu, W.~X. Zhao, and J.-R. Wen, ``Filter-enhanced mlp is all you
  need for sequential recommendation,'' in \emph{Proceedings of the ACM Web
  Conference 2022}, 2022, pp. 2388--2399.

\bibitem{GFNet}
Y.~Rao, W.~Zhao, Z.~Zhu, J.~Lu, and J.~Zhou, ``Global filter networks for image
  classification,'' in \emph{Advances in Neural Information Processing Systems
  34, NeurIPS 2021, December 6-14, 2021, virtual}, 2021, pp. 980--993.

\bibitem{merit_11}
A.~H. Crespo and I.~R. Del~Bosque, ``The influence of the commercial features
  of the internet on the adoption of e-commerce by consumers,''
  \emph{Electronic Commerce Research and Applications}, vol.~9, no.~6, pp.
  562--575, 2010.

\bibitem{merit_19}
C.-L. Hsu and H.-P. Lu, ``Consumer behavior in online game communities: A
  motivational factor perspective,'' \emph{Computers in Human Behavior},
  vol.~23, no.~3, pp. 1642--1659, 2007.

\bibitem{merit_21}
L.~Jin, Y.~Chen, T.~Wang, P.~Hui, and A.~V. Vasilakos, ``Understanding user
  behavior in online social networks: A survey,'' \emph{IEEE communications
  magazine}, vol.~51, no.~9, pp. 144--150, 2013.

\bibitem{autoformer}
H.~Wu, J.~Xu, J.~Wang, and M.~Long, ``Autoformer: Decomposition transformers
  with auto-correlation for long-term series forecasting,'' \emph{Advances in
  Neural Information Processing Systems}, vol.~34, pp. 22\,419--22\,430, 2021.

\bibitem{wiener}
N.~Wiener, ``Generalized harmonic analysis,'' \emph{Acta mathematica}, vol.~55,
  no.~1, pp. 117--258, 1930.

\bibitem{Caser}
J.~Tang and K.~Wang, ``Personalized top-n sequential recommendation via
  convolutional sequence embedding,'' in \emph{Proceedings of the eleventh ACM
  international conference on web search and data mining}, 2018, pp. 565--573.

\bibitem{S3Rec}
K.~Zhou, H.~Wang, W.~X. Zhao, Y.~Zhu, S.~Wang, F.~Zhang, Z.~Wang, and J.-R.
  Wen, ``S3-rec: Self-supervised learning for sequential recommendation with
  mutual information maximization,'' in \emph{Proceedings of the 29th ACM
  International Conference on Information \& Knowledge Management}, 2020, pp.
  1893--1902.

\bibitem{related28}
L.~Wu, S.~Li, C.-J. Hsieh, and J.~Sharpnack, ``Sse-pt: Sequential
  recommendation via personalized transformer,'' in \emph{Fourteenth ACM
  Conference on Recommender Systems}, 2020, pp. 328--337.

\bibitem{related15}
J.~Lin, W.~Pan, and Z.~Ming, ``Fissa: fusing item similarity models with
  self-attention networks for sequential recommendation,'' in \emph{Fourteenth
  ACM Conference on Recommender Systems}, 2020, pp. 130--139.

\bibitem{related7}
Y.~He, Y.~Zhang, W.~Liu, and J.~Caverlee, ``Consistency-aware recommendation
  for user-generated item list continuation,'' in \emph{Proceedings of the 13th
  International Conference on Web Search and Data Mining}, 2020, pp. 250--258.

\bibitem{related5}
X.~Fan, Z.~Liu, J.~Lian, W.~X. Zhao, X.~Xie, and J.-R. Wen, ``Lighter and
  better: low-rank decomposed self-attention networks for next-item
  recommendation,'' in \emph{Proceedings of the 44th International ACM SIGIR
  Conference on Research and Development in Information Retrieval}, 2021, pp.
  1733--1737.

\bibitem{related29}
Q.~Wu, Y.~Gao, X.~Gao, P.~Weng, and G.~Chen, ``Dual sequential prediction
  models linking sequential recommendation and information dissemination,'' in
  \emph{Proceedings of the 25th ACM SIGKDD international conference on
  knowledge discovery \& data mining}, 2019, pp. 447--457.

\bibitem{related18}
Z.~Liu, Z.~Fan, Y.~Wang, and P.~S. Yu, ``Augmenting sequential recommendation
  with pseudo-prior items via reversely pre-training transformer,'' in
  \emph{Proceedings of the 44th international ACM SIGIR conference on Research
  and development in information retrieval}, 2021, pp. 1608--1612.

\bibitem{related3}
Z.~Cui, Y.~Cai, S.~Wu, X.~Ma, and L.~Wang, ``Motif-aware sequential
  recommendation,'' in \emph{Proceedings of the 44th International ACM SIGIR
  Conference on Research and Development in Information Retrieval}, 2021, pp.
  1738--1742.

\bibitem{STOSA}
Z.~Fan, Z.~Liu, Y.~Wang, A.~Wang, Z.~Nazari, L.~Zheng, H.~Peng, and P.~S. Yu,
  ``Sequential recommendation via stochastic self-attention,'' in
  \emph{Proceedings of the ACM Web Conference 2022}, 2022, pp. 2036--2047.

\bibitem{related35}
I.~Pitas, \emph{Digital image processing algorithms and applications}.\hskip
  1em plus 0.5em minus 0.4em\relax John Wiley \& Sons, 2000.

\bibitem{related1}
G.~A. Baxes, \emph{Digital image processing: principles and
  applications}.\hskip 1em plus 0.5em minus 0.4em\relax John Wiley \& Sons,
  Inc., 1994.

\bibitem{SIGIR2022}
S.~Peng, K.~Sugiyama, and T.~Mine, ``Less is more: Reweighting important
  spectral graph features for recommendation,'' in \emph{Proceedings of the
  45th International ACM SIGIR Conference on Research and Development in
  Information Retrieval}, 2022, pp. 1273--1282.

\bibitem{graph}
M.~Cheung, J.~Shi, O.~Wright, L.~Y. Jiang, X.~Liu, and J.~M. Moura, ``Graph
  signal processing and deep learning: Convolution, pooling, and topology,''
  \emph{IEEE Signal Processing Magazine}, vol.~37, no.~6, pp. 139--149, 2020.

\bibitem{learninginFD}
K.~Xu, M.~Qin, F.~Sun, Y.~Wang, Y.-K. Chen, and F.~Ren, ``Learning in the
  frequency domain,'' in \emph{Proceedings of the IEEE/CVF Conference on
  Computer Vision and Pattern Recognition}, 2020, pp. 1740--1749.

\bibitem{fastfourierconvolution}
L.~Chi, B.~Jiang, and Y.~Mu, ``Fast fourier convolution,'' \emph{Advances in
  Neural Information Processing Systems}, vol.~33, pp. 4479--4488, 2020.

\bibitem{resolution}
R.~Suvorov, E.~Logacheva, A.~Mashikhin, A.~Remizova, A.~Ashukha, A.~Silvestrov,
  N.~Kong, H.~Goka, K.~Park, and V.~Lempitsky, ``Resolution-robust large mask
  inpainting with fourier convolutions,'' in \emph{Proceedings of the IEEE/CVF
  Winter Conference on Applications of Computer Vision}, 2022, pp. 2149--2159.

\bibitem{FFC-SE}
I.~Shchekotov, P.~Andreev, O.~Ivanov, A.~Alanov, and D.~Vetrov, ``Ffc-se: Fast
  fourier convolution for speech enhancement,'' \emph{arXiv preprint
  arXiv:2204.03042}, 2022.

\bibitem{NLP1}
A.~Tamkin, D.~Jurafsky, and N.~Goodman, ``Language through a prism: A spectral
  approach for multiscale language representations,'' \emph{Advances in Neural
  Information Processing Systems}, vol.~33, pp. 5492--5504, 2020.

\bibitem{FNET}
J.~Lee-Thorp, J.~Ainslie, I.~Eckstein, and S.~Ontanon, ``Fnet: Mixing tokens
  with fourier transforms,'' \emph{arXiv preprint arXiv:2105.03824}, 2021.

\bibitem{FEDformer}
T.~Zhou, Z.~Ma, Q.~Wen, X.~Wang, L.~Sun, and R.~Jin, ``Fedformer: Frequency
  enhanced decomposed transformer for long-term series forecasting,'' in
  \emph{International Conference on Machine Learning, {ICML} 2022, 17-23 July
  2022, Baltimore, Maryland, {USA}}, ser. Proceedings of Machine Learning
  Research, vol. 162, 2022, pp. 27\,268--27\,286.

\bibitem{FNO}
Z.~Li, N.~Kovachki, K.~Azizzadenesheli, B.~Liu, K.~Bhattacharya, A.~Stuart, and
  A.~Anandkumar, ``Fourier neural operator for parametric partial differential
  equations,'' \emph{arXiv preprint arXiv:2010.08895}, 2020.

\bibitem{AFNO}
J.~Guibas, M.~Mardani, Z.~Li, A.~Tao, A.~Anandkumar, and B.~Catanzaro,
  ``Adaptive fourier neural operators: Efficient token mixers for
  transformers,'' \emph{arXiv preprint arXiv:2111.13587}, 2021.

\bibitem{UFNO}
G.~Wen, Z.~Li, K.~Azizzadenesheli, A.~Anandkumar, and S.~M. Benson,
  ``U-fno—an enhanced fourier neural operator-based deep-learning model for
  multiphase flow,'' \emph{Advances in Water Resources}, vol. 163, p. 104180,
  2022.

\bibitem{R-drop}
L.~Wu, J.~Li, Y.~Wang, Q.~Meng, T.~Qin, W.~Chen, M.~Zhang, T.-Y. Liu
  \emph{et~al.}, ``R-drop: Regularized dropout for neural networks,''
  \emph{Advances in Neural Information Processing Systems}, vol.~34, pp.
  10\,890--10\,905, 2021.

\bibitem{Timeinterval}
J.~Li, Y.~Wang, and J.~McAuley, ``Time interval aware self-attention for
  sequential recommendation,'' in \emph{Proceedings of the 13th international
  conference on web search and data mining}, 2020, pp. 322--330.

\bibitem{KDD}
W.~Krichene and S.~Rendle, ``On sampled metrics for item recommendation,'' in
  \emph{KDD}, 2020.

\bibitem{BPR_MF}
S.~Rendle, C.~Freudenthaler, Z.~Gantner, and L.~Schmidt-Thieme, ``Bpr: Bayesian
  personalized ranking from implicit feedback,'' \emph{arXiv preprint
  arXiv:1205.2618}, 2012.

\bibitem{DFT1}
L.~R. Rabiner and B.~Gold, ``Theory and application of digital signal
  processing,'' \emph{Englewood Cliffs: Prentice-Hall}, 1975.

\bibitem{DFT2}
S.~S. Soliman and M.~D. Srinath, ``Continuous and discrete signals and
  systems,'' \emph{Englewood Cliffs}, 1990.

\bibitem{cooley}
J.~W. Cooley and J.~W. Tukey, ``An algorithm for the machine calculation of
  complex fourier series,'' \emph{Mathematics of computation}, vol.~19, no.~90,
  pp. 297--301, 1965.

\bibitem{FFTW3}
M.~Frigo and S.~G. Johnson, ``The design and implementation of fftw3,''
  \emph{Proceedings of the IEEE}, vol.~93, no.~2, pp. 216--231, 2005.

\end{thebibliography}

\end{document}